\documentclass[prd,aps,nofootinbib,superscriptaddress,floatfix,preprintnumbers,twocolumn]{revtex4}
\usepackage{amssymb}
\usepackage{mathrsfs}
\usepackage{amsfonts}
\usepackage{amsmath,amssymb,amsfonts}
\usepackage{slashed}
\usepackage{array}
\usepackage{float}
\usepackage{verbatim}
\usepackage{epsfig}
\usepackage{graphicx}
\usepackage{color}
\usepackage[colorlinks=true, linkcolor=blue, citecolor=blue, urlcolor=blue]{hyperref}
\usepackage[dvipsnames]{xcolor}
\usepackage{multirow}
\usepackage{lipsum}
\usepackage{cleveref}
\usepackage{comment}
\usepackage{physics}
\usepackage{mathtools}
\usepackage[utf8]{inputenc}
\usepackage{ulem}

\hypersetup{
    colorlinks=true,
    linkcolor=blue,
    filecolor=magenta,      
    urlcolor=blue,}

\newcommand{\LambdaQCD}{\Lambda_\text{QCD}}

\DeclareRobustCommand{\Eq}[1]{Eq.~\eqref{eq:#1}}

\DeclareRobustCommand{\fig}[1]{Fig.~\ref{fig:#1}}

\DeclareRobustCommand{\sec}[1]{Sec.~\ref{sec:#1}}

\DeclareRobustCommand{\refcite}[1]{Ref.~\cite{#1}}

\newcommand\bets{\begin{table*}}
\newcommand\eets[1]{\label{tb:#1}\end{table*}}

\begin{document}

\title{Lattice QCD Calculation of $x$-dependent Meson Distribution Amplitudes at Physical Pion Mass with Threshold Logarithm Resummation}

\author{Ian Cloët}
\affiliation{Physics Division, Argonne National Laboratory, Lemont, IL 60439, USA}

\author{Xiang Gao}
\affiliation{Physics Division, Argonne National Laboratory, Lemont, IL 60439, USA}

\author{Swagato Mukherjee}
\affiliation{Physics Department, Brookhaven National Laboratory, Upton, New York 11973, USA}

\author{Sergey Syritsyn}
\affiliation{Physics Department, Brookhaven National Laboratory, Upton, New York 11973, USA}

\author{ Nikhil Karthik}
\affiliation{American Physical Society, Hauppauge, New York 11788, USA}
\affiliation{Department of Physics, Florida International University, Miami, FL 33199, USA}

\author{Peter Petreczky}
\affiliation{Physics Department, Brookhaven National Laboratory, Upton, New York 11973, USA}

\author{Rui Zhang}
\affiliation{Physics Division, Argonne National Laboratory, Lemont, IL 60439, USA}

\author{Yong Zhao}
\affiliation{Physics Division, Argonne National Laboratory, Lemont, IL 60439, USA}

\begin{abstract}
We present a lattice quantum chromodynamics (QCD) calculation of the $x$-dependent pion and kaon distribution amplitudes (DA) in the framework of large momentum effective theory. This calculation is performed on a fine lattice of $a=0.076$~fm at physical pion mass, with the pion boosted to $1.8$~GeV and kaon boosted to $2.3$~GeV. We renormalize the matrix elements in the hybrid scheme and match to $\overline{\rm MS}$ with a subtraction of the leading renormalon in the Wilson-line mass. The perturbative matching is improved by resumming the large logarithms related to the small quark and gluon momenta in the soft-gluon limit. After resummation, we demonstrate that we are able to calculate a range of $x\in[x_0,1-x_0]$ with $x_0=0.25$ for pion and $x_0=0.2$ for kaon with theoretical systematic errors under control. The kaon DA is shown to be slighted skewed, and narrower than pion DA. Although the $x$-dependence cannot be direct calculated beyond these ranges, we estimate higher moments of the pion and kaon DAs {by complementing} our calculation with short-distance factorization.
\end{abstract}

\maketitle

\section{Introduction}

Pseudoscalar meson distribution amplitudes (DAs) $\phi_M(x)$ of a meson $M$ with valence quark $q\bar{q}'$ describe the probability amplitude of finding the meson moving on the light-cone in its minimal Fock state, a quark-antiquark pair $|q\bar{q}'\rangle$ each carrying a momentum fraction of $x$ and $1-x$, respectively. 
They are important universal inputs to hard exclusive processes and form factors~\cite{Beneke:1999br,Beneke:2001ev,Collins:1996fb,CELLO:1990klc} at large momentum transfer $Q^2\gg \Lambda_{\rm QCD}^2$. Especially, the kaon DA gains particular interest because of its relevance to CP violating processes in heavy meson decays, such as $B\to\pi K$ and $D\to K \bar{K}$ ~\cite{Stewart:2003gt,Li:2009wba,Li:2012cfa}, which are important probes to new physics beyond Standard Model. So far, the $x$-dependence of meson DAs are only weakly constrained by experimental measurements of form factors~\cite{CELLO:1990klc,CLEO:1997fho,BaBar:2009rrj,Belle-II:2018jsg}, thus many model-dependent theoretical calculations are suggested, providing very different shapes with different model assumptions~\cite{Chang:2013pq,Shi:2014uwa,deMelo:2015yxk}. A direct $x$-dependence calculation from first principle methods, such as lattice QCD, could improve our understanding of the meson structures.

The meson DAs can be defined as lightlike correlations,
\begin{align}\label{eq.light-coneDA}
{if_M}\phi_M(x,\mu)&= \int\frac{d \eta^-}{2\pi}e^{ixP^+\eta^-}
\times\nonumber\\
&\langle{0}|{\overline{q}'(0)\gamma_5\gamma_+ W(0,\eta^-)q(\eta^-)}|{M(P)}\rangle ,\nonumber
\end{align}
where {$f_M$ is the meson's decay constant}, $W(0,\eta^-)=\hat{\mathcal{P}}\exp\left[-ig\int^{\eta^-}_0\!\!ds\,n_{\mu}A^{\mu}(ns)\right]$ is the Wilson line between the two light-cone coordinates $0$ and $\eta^-=(\eta^0+\eta^3)/\sqrt{2}$ with $\hat{\mathcal{P}}$ denoting the path-ordering operator. The real-time dependence in the light-cone DA definition makes it difficult to simulate directly on a Euclidean lattice. Thus different approaches have been proposed to extract partial information of meson DAs with lattice calculations, including the calculation of their lowest moments $\langle (2x-1)^n\rangle$ from local twist-2 operators~\cite{Kronfeld:1984zv,DelDebbio:2002mq,Braun:2006dg,Arthur:2010xf,Bali:2017ude,RQCD:2019osh} through the operator product expansion (OPE) of the pion correlators, the calculation of non-local correlations that can be related to the meson DAs through OPE or short distance factorization (SDF)~\cite{Braun:2007wv,Braun:2015axa,Bali:2018spj,Detmold:2005gg,Detmold:2021qln,Gao:2022vyh,Blossier:2024wyx}, and the large momentum effective theory (LaMET)~\cite{Ji:2013dva,Ji:2014gla,Ji:2020ect,Zhang:2017bzy,Zhang:2017zfe,Zhang:2020gaj,Hua:2020gnw,LatticeParton:2022zqc,Holligan:2023rex,Baker:2024zcd} that directly calculates the $x$-dependence from a momentum-space factorization. The local twist-2 operator calculation and the SDF approach in principle allows us to obtain a few lowest-order meson DA moments, but the $x$-dependence could only be obtained by fitting with some model assumptions of the shape. On the other hand, LaMET relates a class of observables at finite hadron momentum,  named quasi-distributions, to the light-cone distribution through a factorization in momentum space, with power corrections suppressed by the power of parton momentum $xP_z$ or $(1-x)P_z$, thus provides an approach to calculate the $x$-dependence in a moderate range of $x$ with well-controlled theoretical systematic errors. 

So far, various lattice artifacts and theoretical systematics have been studied in the LaMET calculation of the meson DAs since the first attempts~\cite{Zhang:2017bzy,Zhang:2017zfe}. The theoretical efforts include the lattice renormalization, improvement of the power accuracy, and the inclusion of higher-order effects in perturbation theory. The renormalization of the linear divergence in the quasi-DA correlators~\cite{Chen:2016fxx,Ji:2017oey,Green:2017xeu,Ishikawa:2017faj} uses regularization-independent momentum subtraction (RI/MOM) ~\cite{Constantinou:2017sej,Alexandrou_2017,Chen:2017mzz,Stewart:2017tvs} scheme or the ratio ~\cite{Orginos:2017kos,Fan:2020nzz} scheme in the early stage, which was later improved with the hybrid renormalization scheme~\cite{Ji:2020brr,LatticePartonCollaborationLPC:2021xdx,Gao:2021dbh}. The power corrections resulting from the  linear renormalon ambiguity of order $\LambdaQCD$ in renormalization and perturbative matching is resolved with the leading-renormalon resummation (LRR) method~\cite{Holligan:2023rex,Zhang:2023bxs}, thus the power accuracy is improved to sub-leading order. The high-order effects are examined by including the calculation of next-to-leading order (NLO) matching kernel for quasi-DA~\cite{Zhang:2017bzy,Zhang:2017zfe,Bali:2017gfr,Xu:2018mpf,Liu:2018tox}, and the renormalization group resummation (RGR)~\cite{Gao:2021hxl,Su:2022fiu} and threshold resummation~\cite{Gao:2021hxl,Ji:2023pba,Liu:2023onm,Baker:2024zcd} that resum the logarithm to further include higher-order logarithm contributions. Since the power expansion parameter is related to the physical scale $xP_z$ or $(1-x)P_z$, the LaMET expansion cannot be applied to endpoint regions $x\to\{0,1\}$.
Beyond the range that LaMET calculates, it has also been proposed to constrain the endpoint regions with the global information of DA, such as the lower moments or short distance correlations, extracted from SDF~\cite{Ji:2022ezo,Holligan:2023rex}.
Numerically, the first continuum extrapolation of the DA was carried out in Ref.~\cite{Zhang:2020gaj} with the RI/MOM scheme at heavier-than-physical pion masses, succeeded by a continuum extrapolation at physical pion mass with NLO hybrid-scheme renormalization and matching, but without LRR~\cite{LatticeParton:2022zqc}. The LRR method was first implemented in Ref.~\cite{Holligan:2023rex} with NLO matching and RGR. Meanwhile, the meson DA moments have also been calculated using the SDF approach at NLO accuracy~\cite{Bali:2018spj,Gao:2022vyh}.
Recently, there is new progress to develop the threshold resummation for the LaMET calculation of pion-DA on a domain-wall ensemble that preserves chiral symmetry~\cite{Baker:2024zcd}.

In this work, we present the $x$-dependence calculation of pion and kaon DAs on a lattice ensemble with physical pion mass and lattice spacing $a= 0.076$ fm, by boosting the pion momentum up to $1.8$~GeV and the kaon momentum up to $2.3$~GeV. We renormalize the bare meson quasi-DA matrix elements with the hybrid scheme, and then match them to the continuum $\overline{\rm MS}$ scheme with LRR. 
After Fourier transforming to the $x$-space, we match the quasi-DA to the light-cone with next-to-next-to-leading logarithmic (NNLL) threshold resummation and NLO matching, which provides a reliable calculation in the range $x\in[0.25,0.75]$ for pion and $x\in[0.2,0.8]$ for kaon. 
Finally, we utilize the short distance correlations along with the already determined mid-$x$ distribution, to better constrain the distribution in the endpoint region, thus provide us a rough estimate for higher moments of DA. 

This work is organized as follows. In Sec.~\ref{sec:lat_setup}, we present our lattice setup of the calculation, and show the extraction of the bare matrix elements. In Sec.~\ref{sec:quasi_da}, we renormalize the bare matrix elements in the hybrid scheme with LRR and obtain $x$-dependent quasi-DAs. In Sec.~\ref{sec:matching}, we apply the threshold-resummed perturbative inverse matching to extract the light-cone DA in a range of $x\in[x_0,1-x_0]$, and provide a rough estimate for the higher moments of pion and kaon DAs by modeling the endpoint regions using complementarity with SDF. Finally, we conclude in Sec.~\ref{sec:conlusion}.

\section{Numerical setup}
\label{sec:lat_setup}

We use a 2+1 flavor Highly Improved Staggered Quark (HISQ)~\cite{Follana:2006rc} gauge ensemble generated by the HotQCD collaboration~\cite{Bazavov:2019www} with lattice size $L_s\times L_t=64^3\times64$ and lattice spacing of a = 0.076 fm. The quark masses in the sea are both at the physical point. We use the Wilson-Clover action for the valence sector, and the clover coefficients are set to be $c_{sw}$ = 1.0372~\cite{Gao:2022iex} using the averaged plaquette after 1-step HYP smearing~\cite{Hasenfratz:2001hp}. The Wilson-Clover quark mass are tuned so that the pion and kaon mass are 140(1) MeV and 498(1) MeV, respectively. The calculations used QUDA multigrid algorithm~\cite{Brannick:2007ue, Clark:2009wm, Babich:2011np, Clark:2016rdz} for the Wilson-Dirac operator inversions to get the quark propagators. {350 configurations are used for the calculation of kaon DA, combined with All Mode Averaging (AMA) technique~\cite{Shintani:2014vja} applied to increase the statistics. On each configuration, we measured 8 exact and 256 sloppy sources for large momenta $n_z\geq4$ and 4 exact and 128 sloppy sources for smaller momenta, respectively.} The pion correlators on this ensemble have already been generated and analyzed in a previous work~\cite{Gao:2022vyh}, so we only present the analysis of kaon raw data in this section.

In order to derive the bare matrix elements of ground state, we need to compute the two-point functions to extract the energy spectrum and get the overlap amplitudes,
\begin{align}
C_{\rm KK}(\mathbf{P},t_s)=\left\langle [K_{S}(\mathbf{P},t_s)] [K_{S}(\mathbf{P},0)]^\dagger\right\rangle,
\end{align}
where $K_{S}$ are the kaon interpolators, namely, 
%\YZ{confirm the notation for kaon interpolators below. $K$ and $\Pi$ has been used interchangeably.}
\begin{align}
\begin{split}
K_S(\mathbf{P},t)=&\sum_{\mathbf{x}}\bar{s}_S(\mathbf{x},t)\gamma_5u_S(\mathbf{x},t)e^{-i\mathbf{P}\cdot\mathbf{x}},\\
\end{split}
\end{align}
in which boosted smearing~\cite{Bali:2016lva, Izubuchi:2019lyk} (denoted by $S$) is applied to have better overlap with ground states and improve the signal of high-momenta states. The Gaussian radius of the light and strange quarks used in this work are $r_l^G$ = 0.59 fm and $r_s^G$ = 0.83 fm. With periodic boundary condition, the hadron momentum in physical unit are $P_z = 2\pi n_z/(L_sa)$ with $n_z$ being an integer.

Similar to \refcite{Gao:2022vyh}, the quasi-DA matrix elements of kaon can be extracted from the equal-time correlators,
\begin{align}
\begin{split}
    C_{\rm{DA}}^\Gamma(z;\textbf{P},t_s)=\left\langle [O_\Gamma(z;\mathbf{P},t_s)]  [K_{S}(\mathbf{P},0)]^\dagger\right\rangle,
\end{split}
\end{align}
with,
\begin{align}
\begin{split}
    &O_\Gamma(z;\mathbf{P},t_s)=\\
    &\sum_{\mathbf{x}} \bar{u}(\mathbf{x},t_s)\Gamma W(\mathbf{x},t_s;\mathbf{x}+\mathbf{z},t_s)s(\mathbf{x}+\mathbf{z},t_s)e^{-i\mathbf{P}\cdot\mathbf{x}},\nonumber
\end{split}
\end{align}
where the quark fields are separated by $\mathbf{z}=(0,0,z)$ and connected by the Wilson line 
$W(\mathbf{x},t_s;\mathbf{x}+\mathbf{z},t_s)$ to keep the gauge invariance. The direction of momentum is chosen to be along with the Wilson line $\mathbf{P}=(0,0,P_z)$. The quasi-DA operator with both $\Gamma=\gamma_5\gamma_3$ and $\gamma_5\gamma_0$ can approach the light-cone DAs under large momentum boost. However, the $\gamma_5\gamma_0$ is not favored due to its mixing under renormalization with the $\sigma_{xy}$ operator from the explicit chiral symmetry breaking of Wilson-type action~\cite{Constantinou:2017sej,Chen:2017mie}. The choice $\gamma_5\gamma_3$ is free of such mixing and we will focus on this choice in the following analysis.

\subsection{Energy spectrum from analysis of $C_{\rm KK}$ correlators}\label{sec:da2pt}

\begin{figure}
\centering
	\includegraphics[width=0.45\textwidth]{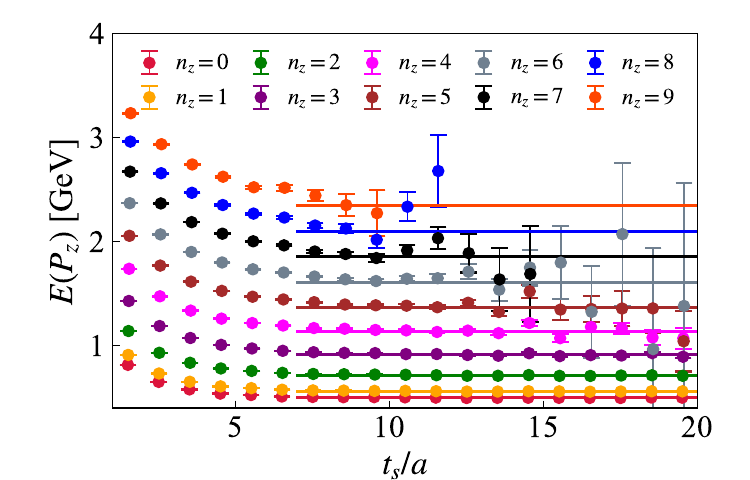}
 \caption{Effective mass of the kaon from the smeared-smeared (SS) correlators, along with the lines representing energies obtained from dispersion relation $E(P_z)=\sqrt{P_z^2+m_K^2}$.\label{fig:effmassHISQ}}
	%\caption{Effective mass of a = 0.076 fm (left panel) ensemble and 0.04 fm (right panels) are shown. \YZ{There is only one panel here, no $a=0.04$ fm ensemble. FIX IT. }The filled and open symbols are corresponding to the SS and SP correlators. \YZ{Define SS and SP.} The lines are calculated from dispersion relation $E(P_z)=\sqrt{P_z^2+m_K^2}$.\label{fig:effmassHISQ}}
\end{figure}

\begin{figure}
\centering
	\includegraphics[width=0.45\textwidth]{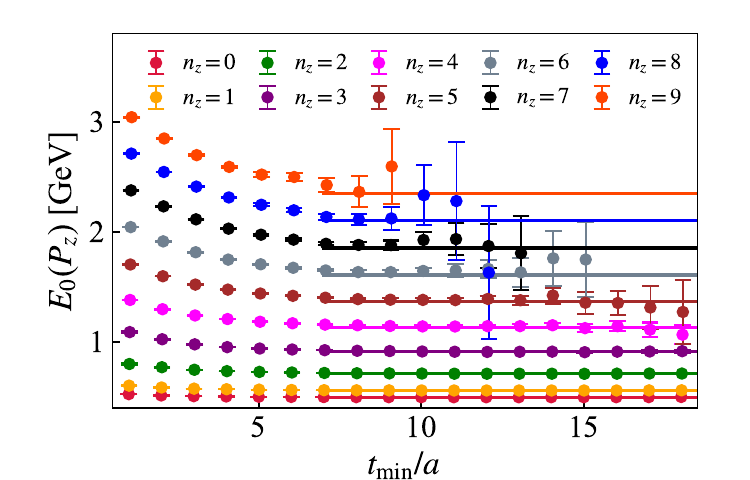}
	\includegraphics[width=0.45\textwidth]{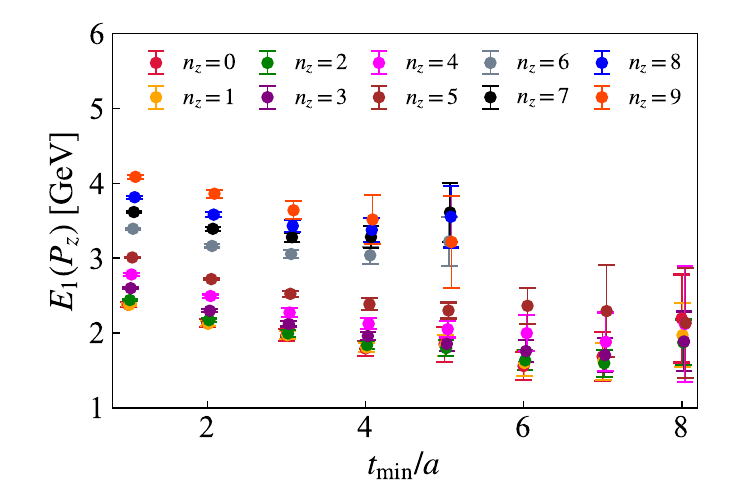}
	\caption{The ground state energy $E_0$ from one-state fit (top) and first excited state energy $E_1$ from two-state fit (bottom) are shown. The colored lines are computed from dispersion relation $E(P_z)=\sqrt{m_K^2+P_z^2}$.\label{fig:da2ptfit}} 
\end{figure}

%Large momentum is the key to extract the distribution amplitudes from the quasi-distributions~\cite{Gao:2022vyh}. In this section, we computed the kaon two-point function with $P_z=2\pi n_z/(N_sa)$ up to 2.29 GeV ($n_z\in[0,9]$). Following the similar strategy to \sec{2pt}, the ground state energy $E_0$ from one-state fit (left panel) and first excited state energy $E_1$ from two-state fit (right panel) are shown in \fig{da2ptfit}. It can be seen that the ground state energy $E_0$ approach plateaus  at large $t_s$ and agree with the colored lines computed from dispersion relation $E(P_z)=\sqrt{m_K^2+P_z^2}$. The first excited state energy $E_1$ show effective plateau when $t_s\gtrsim 4a$, where a two-state form can approximate the spectral decomposition. We will use the overlap amplitudes $A_n=Z_n/\sqrt{2E_n}$ as well the energy levels $E_n$ derived here for the following extraction of DA matrix elements.

The two-point functions of kaon can be decomposed as,
\begin{align}\label{eq:c2ptsp}
\begin{split}
    C_{\rm KK}(\mathbf{P},t_s)&=\sum_{n=0}^{N_{\rm state}-1} |\langle \Omega|K_S|n;\mathbf{P}\rangle|^2 (e^{-E_n t_s}+e^{-E_n (aL_t-t_s)}),\\
&=
    \sum_{n=0}^{N_{\rm state}-1} \frac{|Z_n|^2}{2E_n} (e^{-E_n t_s}+e^{-E_n (aL_t-t_s)}),
    \end{split}
\end{align}
with $E_{n}$ representing the energy levels and $A_n=Z_n/\sqrt{2E_n}$ being the kaon overlap amplitude $\langle\Omega|K_S|n;\mathbf{P}\rangle$. By truncating at $N_{\rm state}=1$, one can define the effective mass of kaon using adjacent time slices as shown in \fig{effmassHISQ}. As one can see, the effective masses reach the plateaus around $t_{s} \gtrsim 8a$, in which the plateau of $P_z=0$ around 497 MeV is the physical kaon mass. The plateaus of non-zero momentum all agree with the solid lines evaluated from the dispersion relation $E_0(P_z)=\sqrt{P_z^2+m_K^2}$, suggesting the energy spectrum is dominated by the kaon ground state after $t_{s} \sim 8a$.

We then perform the $N$-state fit %\YZ{What is the value of $N_{\rm state}$ in Figure 1?}\RZ{No fit in Fig.1} 
for the two-point functions of $t_s$ in range $[t_{\textup{min}},32a]$. The results are shown in \fig{da2ptfit}. The $E_0$ from one-state fit (upper panel) show similar behavior like \fig{effmassHISQ} but more stable because of combining multiple $t_s$. It is clear that the ground state energy $E_0$ follows the dispersion relation $E_0(P_z)=\sqrt{P_z^2+m_K^2}$ well as also observed in ~\refcite{Ding:2024lfj}. To further extract the first excited state $E_1$, we use two-state fit and fix $E_0$ with the dispersion relation. The results are shown in the lower panel of \fig{da2ptfit}. Since the SS correlators are tuned to have better overlap with ground state, the signal of $E_1$ is much worse. Notably, the $E_1$ determined from two-state fit reaches the plateau around $t_{\textup{min}}\sim 4a$, suggesting the energy spectrum can be well described by the two-state model from there. We will use the $E_n$ and $A_n$ in this section for the extraction of bare matrix elements of quasi-DA.

\subsection{Bare matrix elements}

\begin{figure*}
\centering
	\includegraphics[width=0.45\textwidth]{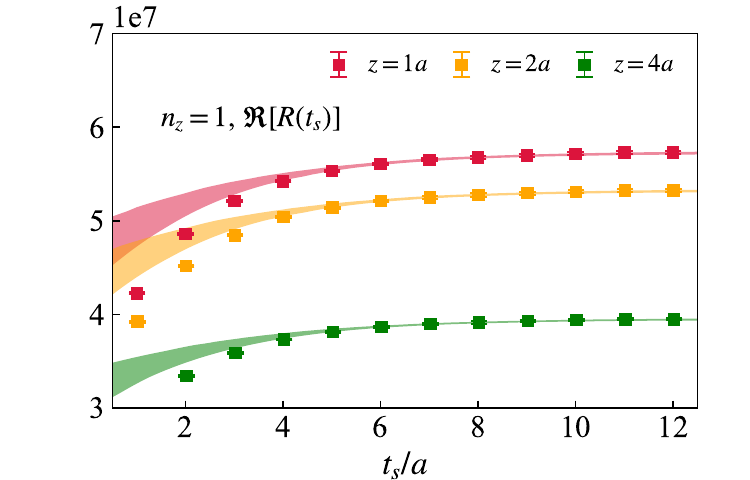}
	\includegraphics[width=0.45\textwidth]{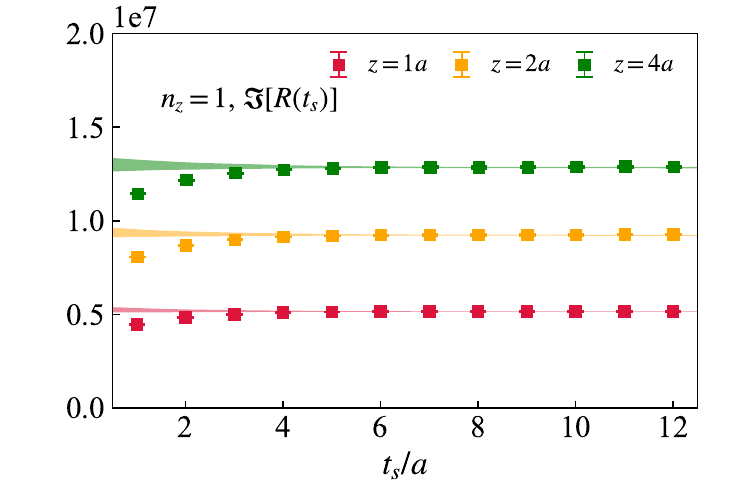}
	\includegraphics[width=0.45\textwidth]{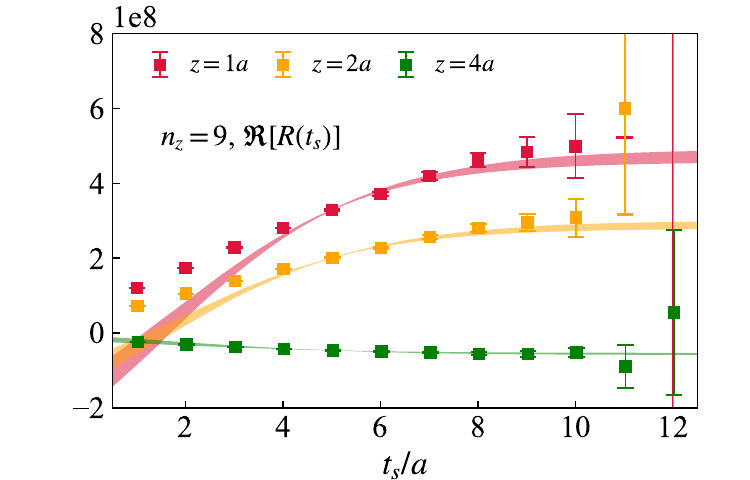}
	\includegraphics[width=0.45\textwidth]{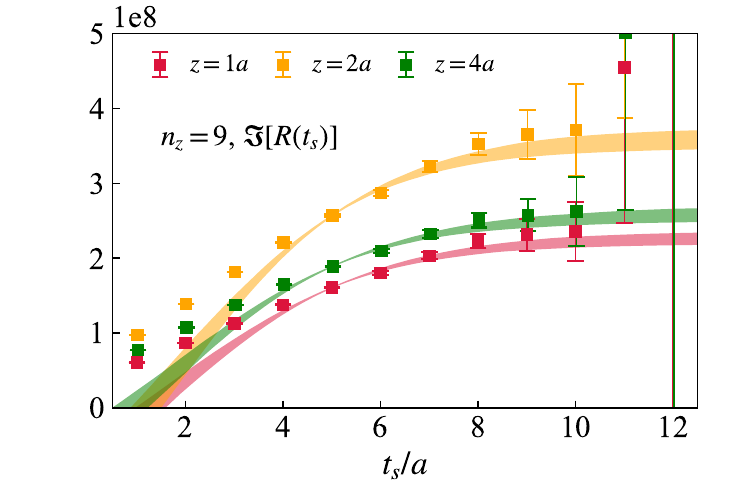}
	\caption{The ratio $R^{\gamma_5\gamma_3}(t_s)=C_{\rm DA}^{\gamma_5\gamma_3}/C_{\rm 2p}$ for momentum $n_z=\{1,9\}$ at $z=\{1a,2a,4a\}$ are shown as the data points. The + panels are for real parts, while the right panels are for the imaginary parts. The bands are results from two-state fit.\label{fig:daRatio}} 
\end{figure*}

\begin{figure}
\centering
	\includegraphics[width=0.45\textwidth]{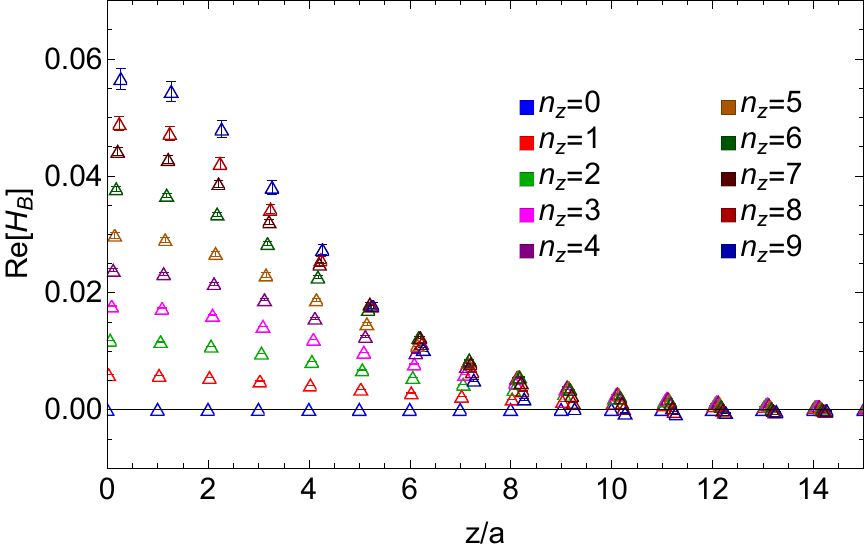}
	\includegraphics[width=0.45\textwidth]{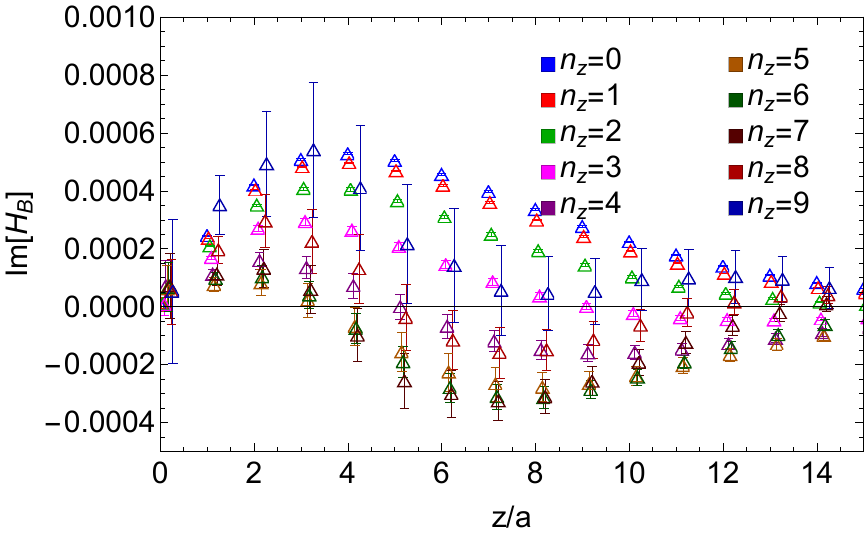}
	\caption{The bare kaon matrix elements $\langle\Omega| O_\Gamma(z)|0;\mathbf{P}\rangle$ for $n_z\in[0,9]$ are shown as a function of $z$.\label{fig:dabm}} 
\end{figure}

The bare quasi-DA matrix elements has spectral decomposition,
%\NK{The DA two-point function is symmetric for $\gamma_3\gamma_5$ and antisymmetric for $\gamma_0\gamma_5$. So there should be a $\pm$ below. }
\begin{align}\label{eq:DAsp}
    C_{\rm DA}^{\gamma_5\gamma_3}(z;\mathbf{P},t_s)=&\sum_{n=0}^{N_{\rm state}-1} \langle\Omega| O_{\gamma_5\gamma_3}(z)|n;\mathbf{P} \rangle\langle n;\mathbf{P}|K_S|\Omega\rangle \nonumber\\
    &\times(e^{-E_n t_s}+e^{-E_n (aL_t-t_s)}),\nonumber\\
=&
    \sum_{n=0}^{N_{\rm state}-1} \frac{Z_n}{2E_n}\langle\Omega| O_{\gamma_5\gamma_3}(z)|n;\mathbf{P} \rangle\nonumber\\& \times(e^{-E_n t_s}+e^{-E_n (aL_t-t_s)}),
\end{align}
where both the $Z_n$ and $E_n$ are the same as the two-point functions defined in \Eq{c2ptsp} and analyzed in \sec{da2pt}. Taking the advantage of correlations between $C_{\rm KK}$ and $C_{\rm DA}$, we construct the ratio,
\begin{align}
R^{\gamma_5\gamma_3}(t_s)=\frac{C_{\rm{DA}}^{\gamma_5\gamma_3}(z;\textbf{P},t_s)}{C_{\rm KK}(\textbf{P},t_s)}.
\end{align}
In \fig{daRatio} we show the ratios $R^{\gamma_5\gamma_3}(t_s)$ for momentum $n_z=\{1,9\}$ at $z=\{1a,2a,4a\}$ as a function of $t_s$. At large $t_s$, plateaus can be clearly observed. In the $t_s\rightarrow \infty$ limit, the quasi-DA matrix elements of kaon ground state can be derived from $R^{\gamma_5\gamma_3}(t_s)\rightarrow \langle\Omega| O_{\gamma_5\gamma_3}(z)|0;\mathbf{P} \rangle/Z_0$. As has been observed in \sec{da2pt} that a two-state spectral decomposition can approximate the $t_s$ dependence of two-point function beyond $t_{\rm min}=4a$, we apply two-state fit to extract the ground state matrix elements, using the ratios $R^{\gamma_5\gamma_3}(t_s)$ started from $t_s\gtrsim 4a$. The fit results are shown as the bands in \fig{daRatio}, which nicely describe the data. The extracted bare matrix elements $\langle\Omega| O_{\gamma_5\gamma_3}(z)|0;\mathbf{P}\rangle$ for $n_z\in[0,9]$ are summarized in \fig{dabm}.

\begin{figure}[!htbp]
\centering
	\includegraphics[width=0.55\textwidth]{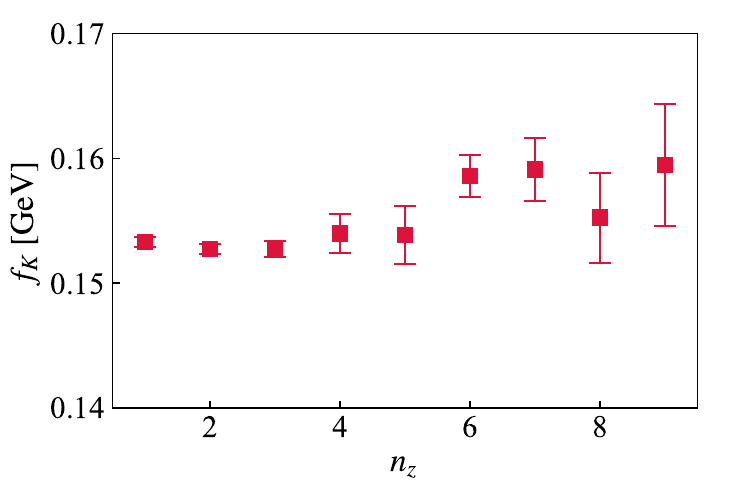}
	\caption{The kaon decay constant $f_K$ is shown as a function of momenta $P_z=2\pi n_z/(N_sa)$.\label{fig:dafK}} 
\end{figure}

% \begin{figure}[!htbp]
% \centering
% 	\includegraphics[width=0.55\textwidth]{{plots/HISQa076_kaon_qDA_bm_P0}.pdf}
% 	\caption{The bare matrix elements $\langle\Omega| O_{\gamma_5\gamma_3}(z)|0;\mathbf{P}\rangle$ for $n_z=0$ are shown as a function of $z$.\label{fig:dabmP0}} 
% \end{figure}

Considering the Lorentz covariance, the matrix elements can be parameterized as~\cite{Bhattacharya:2022aob}, %\YZ{Should the arguments of $\tilde H$ and $\tilde k$ be Lorentz scalars?}
\begin{align}\label{eq:LRZdc}
\begin{split}
	%\langle\Omega| O_0(z)|0;\mathbf{P} \rangle &=P_0 \tilde{h}(z^2,z\cdotP_z),\\
	\langle\Omega| O_{\gamma_5\gamma_\mu}(z)|0;\mathbf{P} \rangle &=P^\mu \tilde{H}(z^2,z\cdot P) + z^\mu m_K^2\tilde{k}(z^2,z\cdot P).
\end{split}
\end{align}
The 2nd term of the right hand side proportional to $z^\mu$ contributes to the matrix elements of $O_{\gamma_5\gamma_3}$ at finite spacial separation $z$, while it is zero for the case of using $O_{\gamma_5\gamma_0}$. In the infinite-momentum limit, the first term $\tilde{H}(z^2,z\cdot P_z)$ is approaching the light-cone DAs, thus is identified as the actual matrix element $\tilde{H}(z,P_z)\equiv \tilde{H}(z^2,z\cdot P_z)$ in our notation, while the $\tilde{k}(z^2,z\cdot P)$ term is suppressed to zero as an power correction. It is worth to mention that, $P^z \tilde{H}(z,P_z)$ is zero when $P_z=0$ for the case of $O_{\gamma_5\gamma_3}$. 

When $z=0$, the kaon decay constant can be extracted from the matrix elements by $\tilde{H}(0,P_z)=f_K/Z_A$ with $Z_A=0.969(1)$ the finite renormalization constant for the axial current operator that has been determined in Ref.~\cite{Gao:2022vyh}. The $f_K$ we got are summarized in \fig{dafK} as a function of momenta $P_z=2\pi n_z/(N_sa)$. As one can observe, the momentum dependence is mild, suggesting the goodness of our extraction. Our determination at $n_z=1$ and 2 are 153.3(4) and 152.7(4) MeV respectively, which are close to the FLAG average for 2+1 flavor QCD $f_K$ = 155.7(7) MeV~\cite{FlavourLatticeAveragingGroupFLAG:2021npn}. The small deviation could come from the lattice artifacts as only one ensemble is used in this calculation. When $z\neq0$, considering the parity and time-reversal symmetry (see appendix in Ref.~\cite{Bhattacharya:2022aob}), the $zm_K^2\tilde{k}(z^2,z\cdot P)$ term can only contribute to the imaginary part of matrix elements and leave the real part unaffected. As shown in \fig{dabm}, the real parts of matrix elements at $P_z=0$ are zero while the imaginary parts are non-zero as expected. In our calculation, to {reduce} the twist-3 contamination from the 2nd term in Eq.~\eqref{eq:LRZdc}, we ignore the $P_z$ dependence in $\tilde{k}(z^2,z\cdot P)$, and define the subtracted matrix elements as
%\NK{Need to put in $O_{\gamma_5\gamma_0}$ explicitly below?}
\begin{align}
    H^B(z,P_z)=\frac{1}{P_\mu}\left(\langle\Omega| O_{\gamma_5\gamma_3}(z)|0;\mathbf{P} \rangle - \langle\Omega| O_{\gamma_5\gamma_3}(z)|0;\mathbf{0} \rangle\right).
\end{align}

\section{Extraction of pion and kaon quasi-DAs}
\label{sec:quasi_da}
\subsection{Moments from SDF}
\begin{figure}[!htbp]
    \centering
    \includegraphics[width=0.45\textwidth]{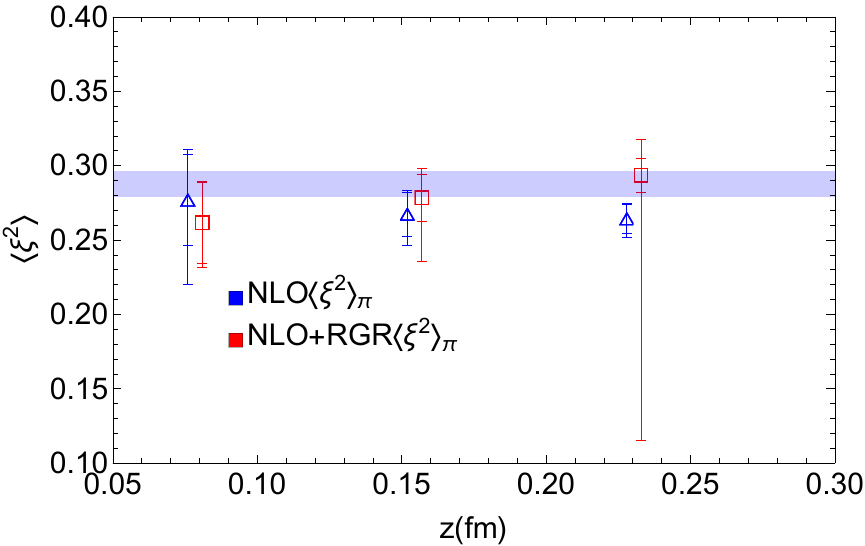}
    \includegraphics[width=0.45\textwidth]{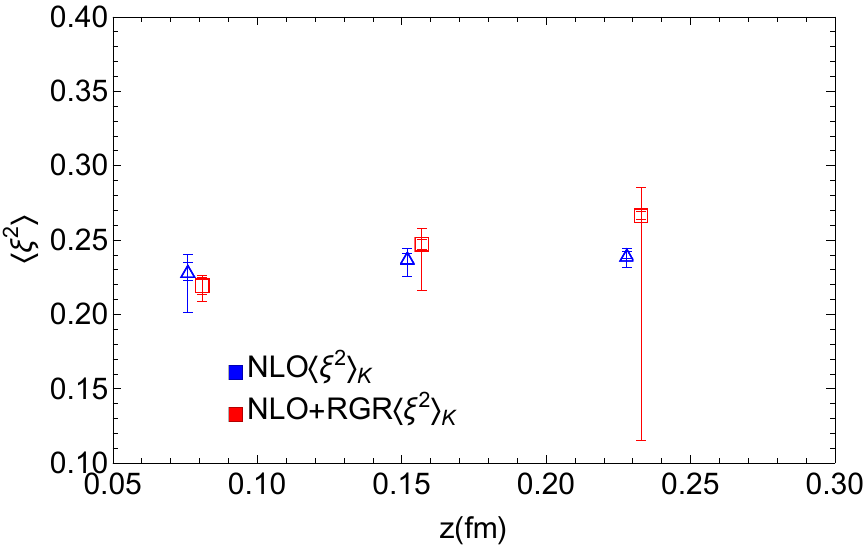}
    \caption{The second moments of pion (top) and kaon (bottom) {at $\mu=2$ GeV} fitted at different $z$ values. The blue band is the pion moment obtained in Ref.~\cite{Gao:2022vyh}.}
    \label{fig:moment_fit}
\end{figure}

The multiplicative renormalizablility of quasi-DA~\cite{Ji:2017oey,Green:2017xeu,Ishikawa:2017faj} allows us to define a UV-finite quantity by taking the double ratio of matrix elements at two different momenta{~\cite{Orginos:2017kos,Fan:2020nzz}},
\begin{align}\label{eq:ratio}
    \mathcal{M}(z,P_1,P_2)&=\lim_{a\to0}\frac{H^B(z,P_2,a)/H^B(0,P_2,a)}{H^B(z,P_1,a)/H^B(0,P_1,a)}\nonumber\\
    &=\frac{H^R(z,P_2,\mu)}{H^R(z,P_1,\mu)},
\end{align}
where we have chosen the normalization $H^B(0,P,a)=H^R(0,P,\mu)=1$. The above ratio is renormalization group invariant, thus is independent of the renormalization scale $\mu$. The ratio can be re-formulated through a short-distance factorization at $z\ll \Lambda_{\rm QCD}^{-1}$ of $H^R(z,P_2,\mu)$ with OPE~\cite{Gao:2022vyh} 
%\YZ{fix the summation. there should be to summations in the following formula}:
\begin{align}\label{eq:ope_da}
    H^R(z,P_z)&=\sum_{n=0}^{\infty}\frac{(\frac{-izP_z}{2})^n}{n!}\sum_{m=0}^{n}C_{nm}(z,\mu)\langle\xi^m\rangle(\mu)
    \nonumber\\&+\mathcal{O}(z^2\Lambda_{\rm QCD}^2),
\end{align}
where $\langle\xi^m\rangle(\mu)=\int_0^1 dx (2x-1)^m\phi(x,\mu)$ are the non-perturbative moments of the light-cone DA, 
and $C_{nm}(z,\mu)$ are the perturbative matching coefficients. Then the ratio $\mathcal{M}(z,P_1,P_2)$ is a function of the moments and known Wilson coefficients when the higher twist correction $\mathcal{O}(z^2\Lambda_{\rm QCD}^2)\ll1$, 
%\YZ{fix the summations}
\begin{align}
    \mathcal{M}(z,P_1,P_2)\approx \frac{\sum_{n=0}^{\infty}\sum_{m=0}^{n}\frac{(\frac{-izP_2}{2})^n}{n!}C_{nm}(z,\mu)\langle\xi^m\rangle(\mu)}{\sum_n^{\infty}\sum_{m=0}^{n}\frac{(\frac{-izP_1}{2})^n}{n!}C_{nm}(z,\mu)\langle\xi^m\rangle(\mu)},
    \label{eq:ratio_ope}
\end{align}
allowing us to extract the moments from these ratios.
Given the better signal-to-noise ratio in the real part of our data, we first extract even moments of $\xi$. Truncating Eq.~\eqref{eq:ratio_ope} at $m,n\leq 4$ and the perturbation series at NLO~\cite{Radyushkin:2019owq,Gao:2022vyh}, we have:
\begin{align}
    &C^{\gamma_i}_{nm}=\delta_{nm}+\frac{\alpha_sC_F}{2\pi}\left(\begin{matrix}
        \frac{3}{2}L+\frac{7}{2}& 0 & 0\\
        -\frac{5}{12}L+\frac{11}{12}&\frac{43}{12}L-\frac{37}{12}  & 0\\
        -\frac{2}{15}L+\frac{1}{4}&-\frac{19}{30}L+\frac{5}{3} &\frac{68}{15}L-\frac{247}{36}
    \end{matrix}\right),
\end{align}
where $L=\ln\frac{z^2\mu^2e^{2\gamma_E}}{4}$. The triangular matrix $C$ is a result of the non-multiplicative renormalization group evolution of the DA moments~\cite{Lepage:1980fj},
\begin{align}
    \label{eq:evo_moment}
&\frac{d}{d\ln \mu^2}\begin{pmatrix}
        1\\
        \langle \xi^2\rangle(\mu)\\
        \langle \xi^4\rangle(\mu)
    \end{pmatrix}
    =\gamma_{nm}\begin{pmatrix}
        1\\
        \langle \xi^2\rangle(\mu)\\
        \langle \xi^4\rangle(\mu)
    \end{pmatrix}
    \\&=
    -\frac{\alpha_s(\mu) C_F}{2\pi}\begin{pmatrix}
        0&0&0\\
        -\frac{5}{12}& \frac{25}{12}&0\\
        -\frac{2}{15}& -\frac{19}{30}& \frac{91}{30}
    \end{pmatrix}\cdot \begin{pmatrix}
        1\\
        \langle \xi^2\rangle(\mu)\\
        \langle \xi^4\rangle(\mu)
    \end{pmatrix}+\mathcal{O}(\alpha_s^2).\nonumber
\end{align}
The off-diagonal part of $\gamma_{nm}$ has been calculated up to  3-loop order using conformal symmetry~\cite{Braun:2017cih}. 
We fit the moments $\{\langle \xi^2\rangle(\mu),\langle \xi^4 \rangle(\mu)\}$ with both NLO and the RG-resummed (RGR) Wilson coefficients. In the latter case, we fit the moments at an initial scale $\langle \xi^2 (\mu_i=2e^{-\gamma_E}z^{-1})\rangle$, where the log terms in $C_{nm}(z,\mu_i)$ vanish, then evolve to $\mu=2$~GeV by solving Eq.~\eqref{eq:evo_moment}. The scale variation is examined by choosing $\mu_i=2ce^{-\gamma_E}z^{-1}$ with $c=\{\sqrt{2},1,1/\sqrt{2}\}$ to estimate the uncertainties from higher-order perturbation theory.
The fitted ratio and the extracted moments are shown in Fig.~\ref{fig:moment_fit}. We show the fitted moments with both statistical and scale variation error bars. Note that the scale variation becomes large when $z$ increases, because the scale $z^{-1}/\sqrt{2}\sim\Lambda_{\rm QCD}$ becomes non-perturbative. The second moment of kaon DA is slightly lower than the moment of pion DA, indicating the kaon DA to be narrower, similar to what has been observed in previous $x$-dependent calculations~\cite{Zhang:2017zfe,Zhang:2020gaj,LatticeParton:2022zqc}.
%\YZ{any reason why it is expected?}

\begin{figure}[!htbp]
    \centering
    \includegraphics[width=0.45\textwidth]{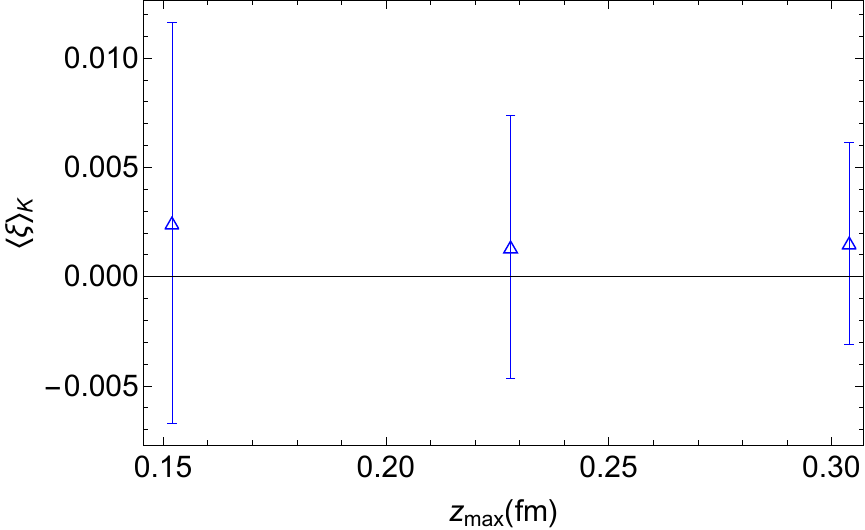}
    \caption{The first moment of kaon at $\mu=2$ GeV fitted at different $z_{\rm max}$ values.}
    \label{fig:moment_fit_odd}
\end{figure}

In principle, the odd moments can also be extracted from the imaginary part of the matrix elements. In our data, we found that the imaginary part are nonvanishing for smaller momenta data, indicating a non-zero first moment of kaon DA. However, for large momenta, the imaginary parts decreases with momentum, and for the largest momentum $\Im[H(z,n_z=9)]$ are consistent with zero at short distance. Thus in principle we cannot describe all data with a universal first moment. Since we focus on the large momentum expansion approach in this work, we show the fit results for the ratio of imaginary part at the largest momentum $n_z=9$ to the real part of $n_z=1$ in the range of  $z\in[a,z_{\rm max}]$ at different $z_{\rm max}$. This results in a first Mellin moment of kaon $\langle \xi\rangle_K(\mu)\approx 0.0013(60)$, corresponding to the first Gegenbauer coefficient $a_1=0.002(10)$, which is consistent with zero, as shown in Fig.~\ref{fig:moment_fit_odd}. {The suppression of the kaon skewness at large momentum may suggest that such skewness originating from the mass difference between light and strange quark is no longer important in the infinite momentum limit, where both partons act just as massless particles.} However, considering that our subtracted imaginary part is still potentially contaminated by the $P_z$ dependence in $z^\mu k(z^2,z\cdot P)$ term in Eq.~\eqref{eq:LRZdc}, and since previous lattice calculations of local twist-2 operators suggest a non-zero kaon first moment~\cite{Braun:2006dg,Donnellan:2007xr,Arthur:2010xf,RQCD:2019osh}, the suppressed imaginary part at large momentum in our data may just be a lattice artifact, and needs further investigation. 
%\YZ{Any explanation? The smallness of the odd moments likely suggests that they mainly originate from the quark mass difference, which is suppressed at $\mu \gg m_s$.}

\subsection{Renormalization}
\begin{figure}[!htbp]
    \centering
    \includegraphics[width=0.33\textwidth]{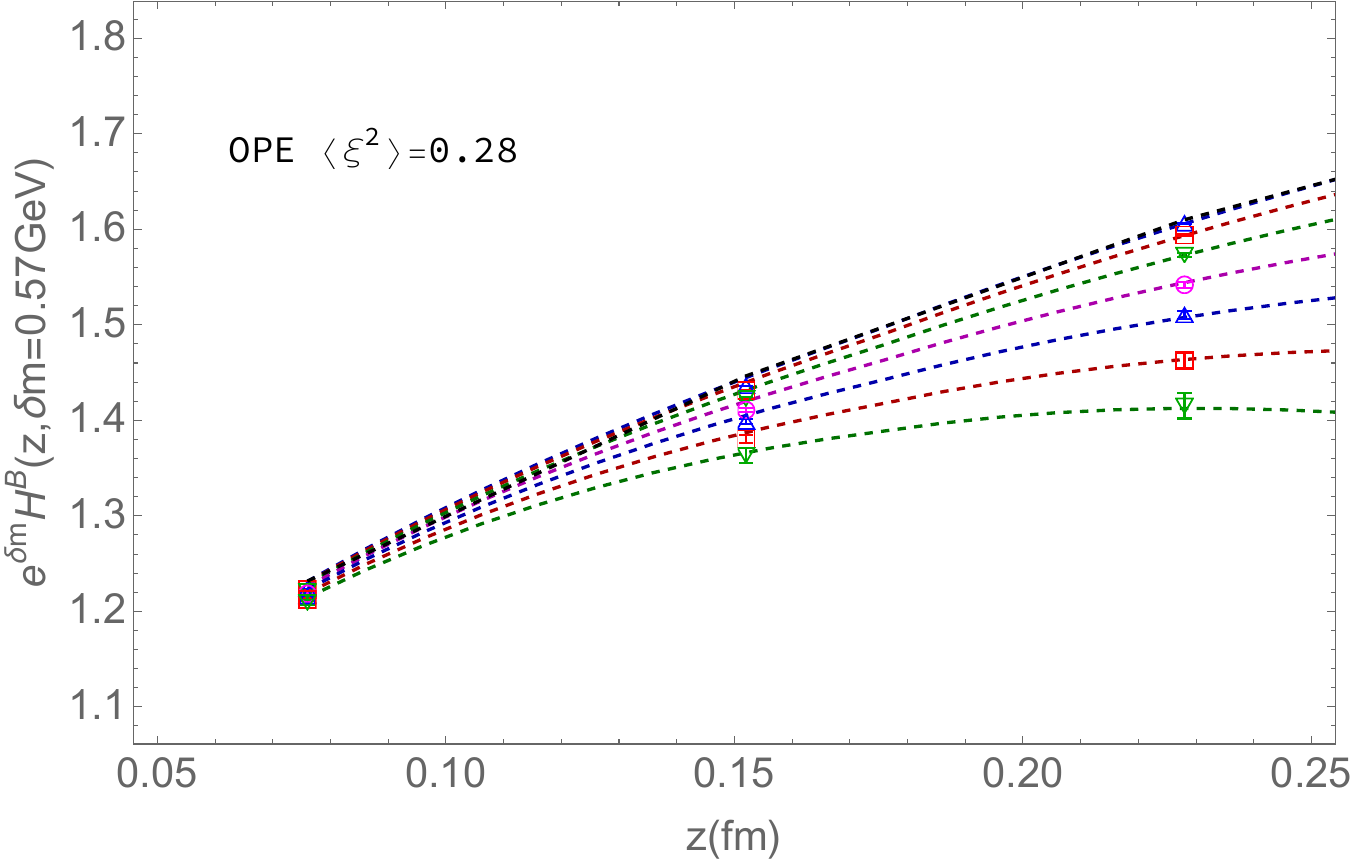}
    \includegraphics[width=0.33\textwidth]{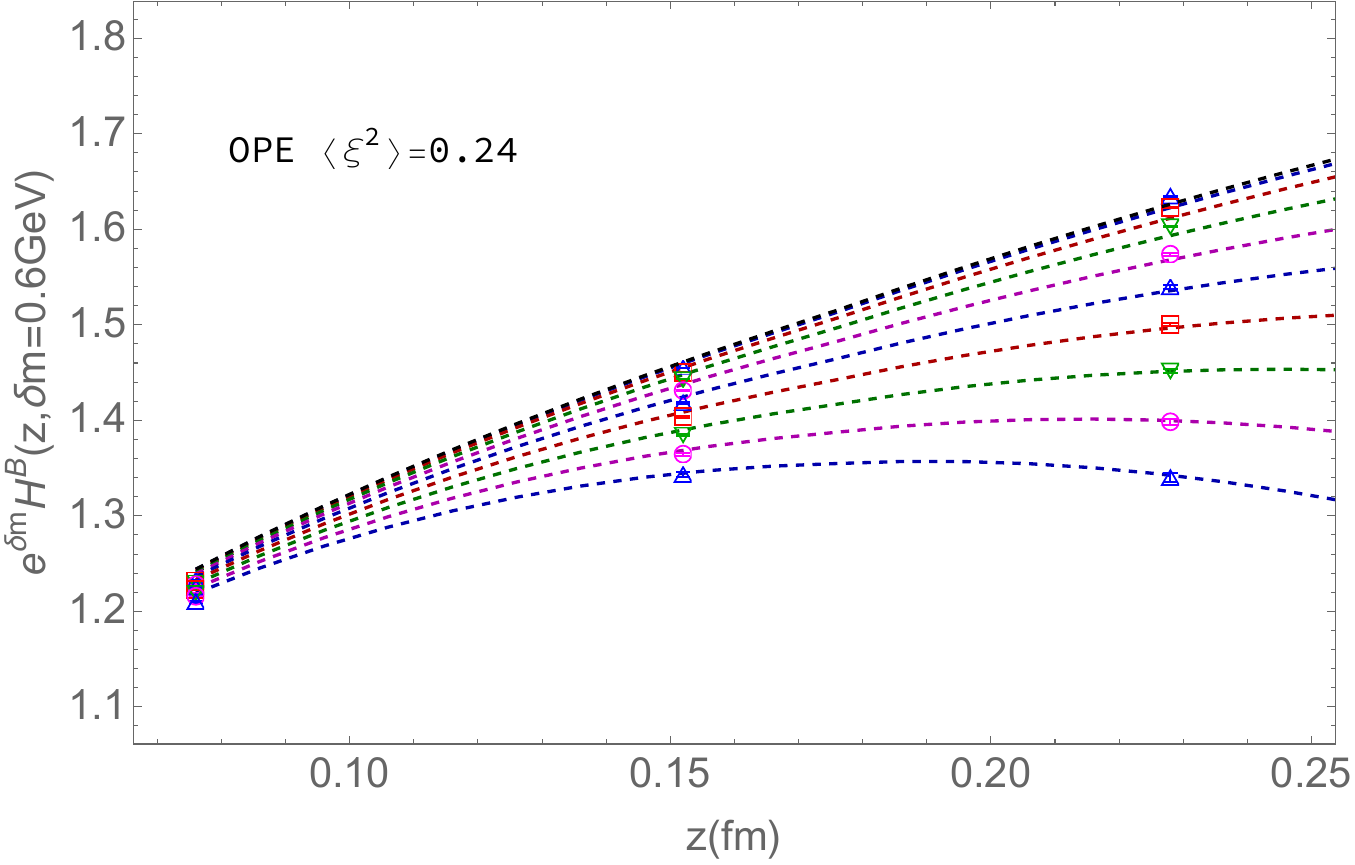}
    \includegraphics[width=0.33\textwidth]{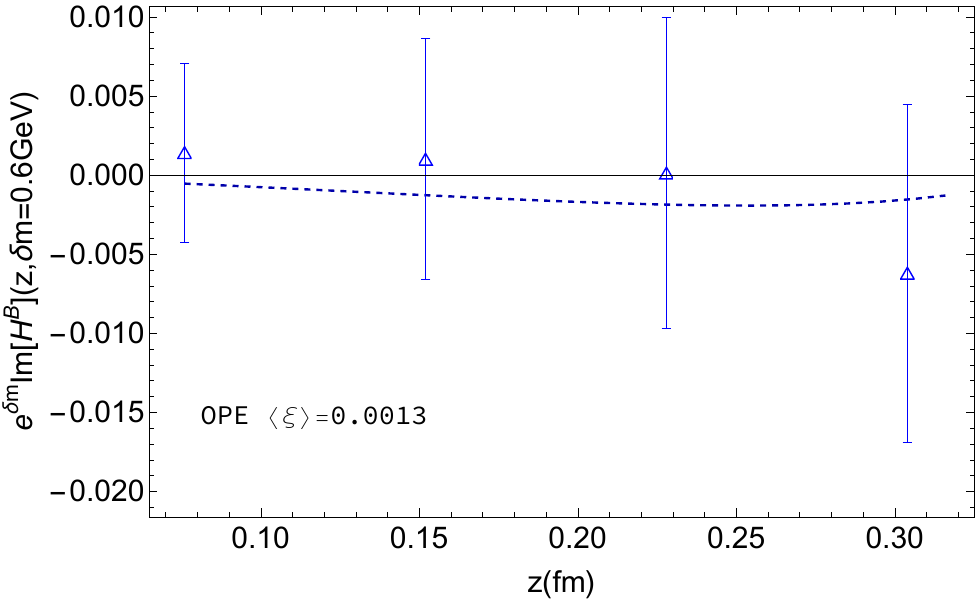}
        \caption{Renormalized matrix element (markers) compared with OPE reconstruction (dashed lines) at short distance for pion (top) and kaon's real (mid) and imaginary part (bottom). Different curves correspond to different momenta.}
    \label{fig:renorm_ope_compare}
\end{figure}

\begin{figure*}[!htbp]
    \centering
    \includegraphics[width=0.45\textwidth]{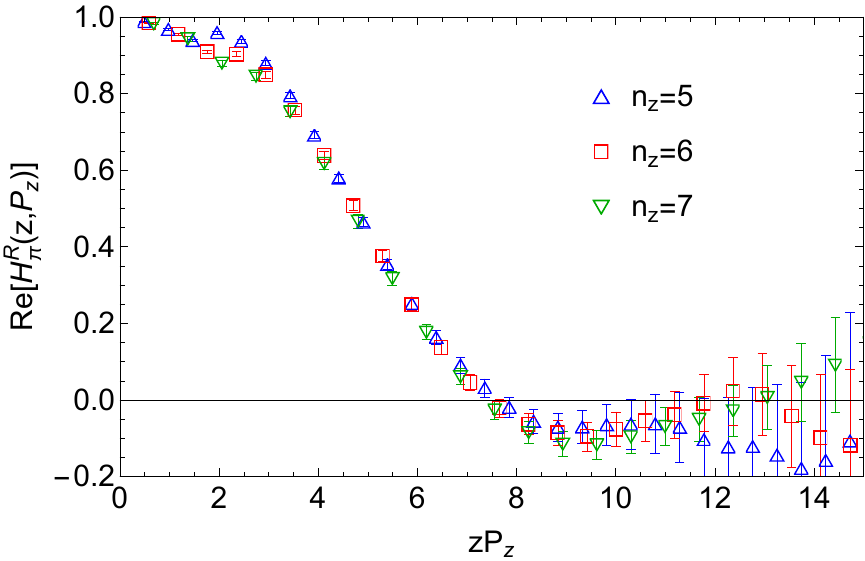}
    \includegraphics[width=0.45\textwidth]{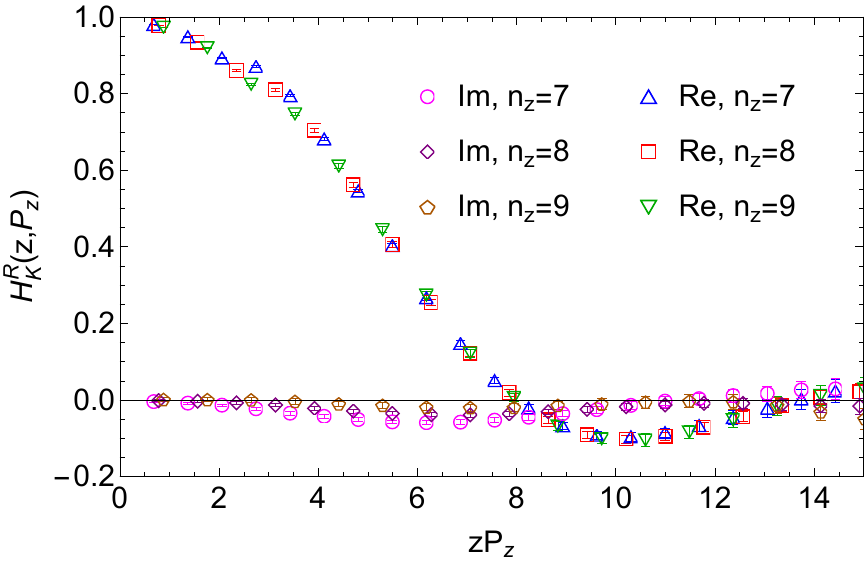}
    \caption{Renormalized matrix elements for pion (left) and kaon (right). At large $zP_z$, the data from different momenta follow a general curve, except for scaling violation in $z^2$ or $1/P_z^2$ which is not distinguishable from the statistical uncertainties. The pion DA matrix elements is purely real because of isospin symmetry, and the kaon DA skewness is shown to be small according to its small imaginary part. }
    \label{fig:renorm_me}
\end{figure*}

The bare matrix elements of meson quasi-DAs contain UV divergences. Especially, besides logarithmic divergences, the spatial Wilson line $W(0,z)$ in the operator has a linearly divergent self energy in the lattice regulator $a^{-1}$~\cite{Chen:2016fxx,Ji:2017oey,Green:2017xeu,Ishikawa:2017faj}. It can be renormalized multiplicatively as~\cite{Ji:2017oey,Green:2017xeu,Ishikawa:2017faj} 
\begin{align}
    H^R(z,P_z)=H^B(z,P_z,a)/Z^R(z,a)
\end{align}
with a renormalization constant $Z^R(z,a)\sim e^{-\delta m(a)|z|}$ containing the same linear divergence in the mass counterterm  $\delta m(a)\sim a^{-1}$. In this work, we used the hybrid renormalization scheme~\cite{Ji:2020brr} that can be perturbatively matched to the $\overline{\rm MS}$ scheme in all regions of $z$. The renormalization constant in the hybrid scheme is defined as
\begin{align}
Z^{\rm hybrid}(z,a)=\left\{
\begin{matrix}
Z^R(z,a)\hfill  &  |z|\leq z_s \\
Z^R(z_s,a)e^{-\delta m(a) (|z|-z_s)} & |z|>z_s
\end{matrix}
\right.
\label{eq:hybrid_renorm}
\end{align}
At short distance, since the correlator of $\gamma_z\gamma_5$ vanishes at $P_z=0$, we cannot use the ratio scheme $Z^R(z,a)=H^B(z,P=0,a)$. Instead, we determine it from the self-renormalization approach~\cite{LatticePartonCollaborationLPC:2021xdx,Holligan:2023rex}, which uses the perturbatively calculated Wilson coefficient
\begin{align}\label{eq:self_renorm}
    Z^R(z,a)=e^{-\delta m(a)|z|}C_0(z,\mu)
\end{align}
for $|z|\leq z_s$. And $\delta m(a)$ is obtained by requiring that the matrix elements after removing the linear divergence $e^{\delta m(a)|z|}H^B(z,P_z,a)$ agree with OPE reconstruction of $H^R(z,P_z,\mu)$ in Eq.~\eqref{eq:ope_da} using the moments fitted from the RG-invariant ratio as inputs, up to a constant conversion factor that converts the lattice scheme to $\overline{\rm MS}$~\cite{Zhang:2023bxs},
\begin{align}
    e^{\delta m(a)|z|}H^B(z,P_z,a)=H^R(z,P_z,\mu) e^{\mathcal{I^{\rm lat}}(a^{-1})-\mathcal{I}(\mu)},
\end{align}
where $\mathcal{I}(\mu)=\int \tfrac{\gamma}{\beta} d\alpha|_{\alpha=\alpha_s(\mu)}$ are the RG-evolution factors that cancels the UV-regulator dependence $\mu$ in the matrix elements.
To ensure the linear power accuracy of the factorization, 
we applied the LRR and RGR-improved Wilson coefficient~\cite{Zhang:2023bxs}.
\begin{align}\label{eq:c0_lrr}
    C_{mn}^{\rm LRR}&(z,z^{-1})=C_{mn}(z,z^{-1})+\delta_{nm}N
    \nonumber\\&\big[-\alpha_s  (1+c_1)
    +\frac{4\pi}{\beta_0}   \int_{\rm 0, PV}^{\infty}du   e^{-\frac{4\pi u}{\alpha_s(\mu)\beta_0}} \nonumber\\& \frac{1}{(1-2u)^{1+b}}\big(1+c_1(1-2u)+...\big)\big],
\end{align}
where $b=\beta_1/2\beta_0^2$ and $c_1=(\beta_1^2-\beta_0\beta_2)/(4b\beta^4_0)$ are from higher orders in the QCD beta function, $N(n_f=3)=0.575$ is the overall strength of the linear renormalon estimated from the quark pole mass correction~\cite{Pineda:2001zq,Bali:2013pla}. 
Taking the moments we fitted from the RG-invariant ratios and tune the value of $\delta m$, we find the matrix elements $e^{\delta m |z| }H^B(z,P_z,O)$ agree well with OPE reconstruction Eq.~\eqref{eq:ope_da} when $\delta m\approx 0.6$~GeV for kaon and $\delta m\approx 0.57$~GeV for pion, as shown in Fig.~\ref{fig:renorm_ope_compare}. 
The consistency with OPE reconstruction suggests that the short-distance data can be well-described by perturbation theory. This justifies our choice of using $C_0(z,\mu)$ calculated from perturbation theory in Eq.~\eqref{eq:self_renorm} to renormalize the short distance matrix elements.
Then we apply it to the hybrid scheme with $z_s=3a$ in Fig.~\ref{fig:renorm_me}. We find the largest three momenta converge to a universal shape, and the imaginary part for kaon is nonvanishing but close to zero, indicating the existence of small skewness.

\subsection{$x$-dependent quasi-DA}
The LaMET factorization is naturally defined in momentum space, which requires the $x$-dependence of quasi-DA to obtain the $x$-dependent light-cone DA.
The extraction of $x$-dependence requires the Fourier transform of the above renormalized correlation functions to the momentum space. Thus in principle we need to know the distribution to arbitrarily large $zP_z$. On the other hand, the noise increases exponentially at large $z$, limiting our calculations to a certain range of $z$ values. An exact Fourier transformation is thus impossible. However, the contribution of the unknown long-tail distribution turns out to be bounded by physical constraints. For example, the Euclidean correlation functions must have a finite correlation length $\lambda_0 \propto P_z$ in the coordinate space, which results in an exponential decay $\exp{[-\lambda/\lambda_0]}$ of the correlation functions at large $\lambda=zP_z$.  We are thus able to make an extrapolation of the long-tail distribution with these constraints to reduce uncertainty from it. 
One example of the long-tail modeling is~\cite{Ji:2020brr},
\begin{align}\label{eq:longtail_form}
    H(\lambda)\xrightarrow{\lambda\to\infty}\left(\frac{c_1e^{-i\lambda/2}}{(-i\lambda)^{d_1}}+\frac{c_2e^{i\lambda/2}}{(i\lambda)^{d_2}}\right)e^{-\lambda/\lambda_0},
\end{align}
{where the parametrization inside the round brackets is motivated from the Regge behavior~\cite{Regge:1959mz} of the light-cone distribution near the endpoint regions.} For symmetric pion DA, $c_1=c_2$ and $d_1=d_2$. For kaon DA, since the imaginary part is too small to be extrapolated, we model only the real part, and directly discrete Fourier transform the imaginary part truncated at the point when it starts to be consistent with zero. The model dependence of the extrapolation is estimated from three different fits of the long tail, by turning off the exponential decay $e^{-\lambda/\lambda_0}$ (labeled with ``$\lambda_0\to\infty$'') in Eq.~\eqref{eq:longtail_form}, or by extrapolating from a different $\lambda_{\rm cut}=z_{\rm cut}P_z$, either from the trough of wave or from a larger $\lambda_{\rm cut}$ where the decaying mode starts to dominate.

\begin{figure*}[!htbp]
    \centering
    \includegraphics[width=0.45\textwidth]{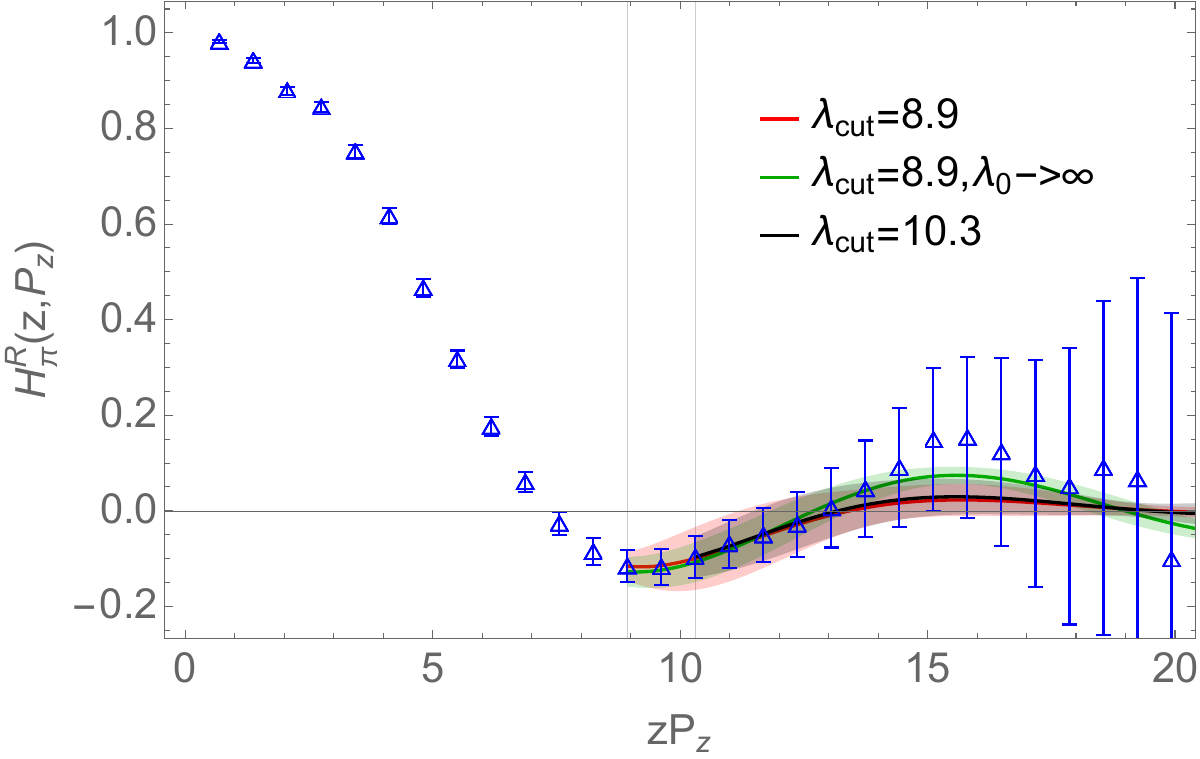}
    \includegraphics[width=0.45\textwidth]{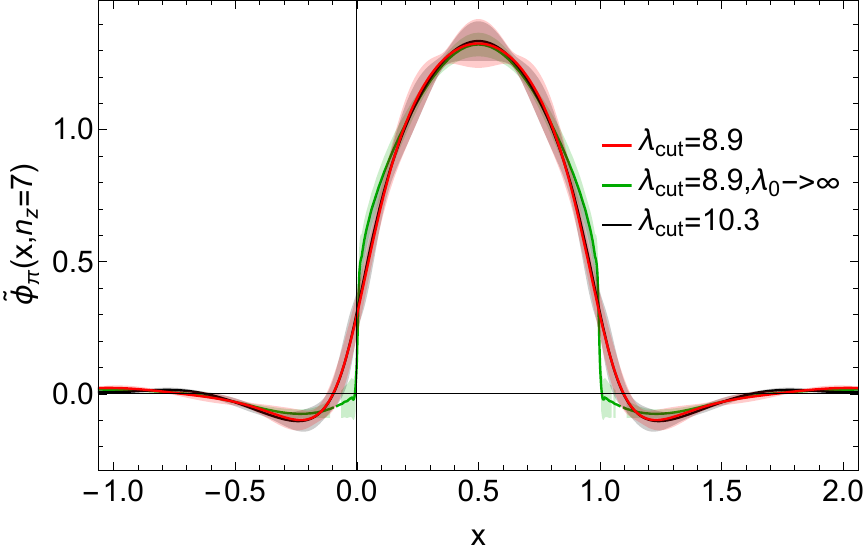}
    \includegraphics[width=0.45\textwidth]{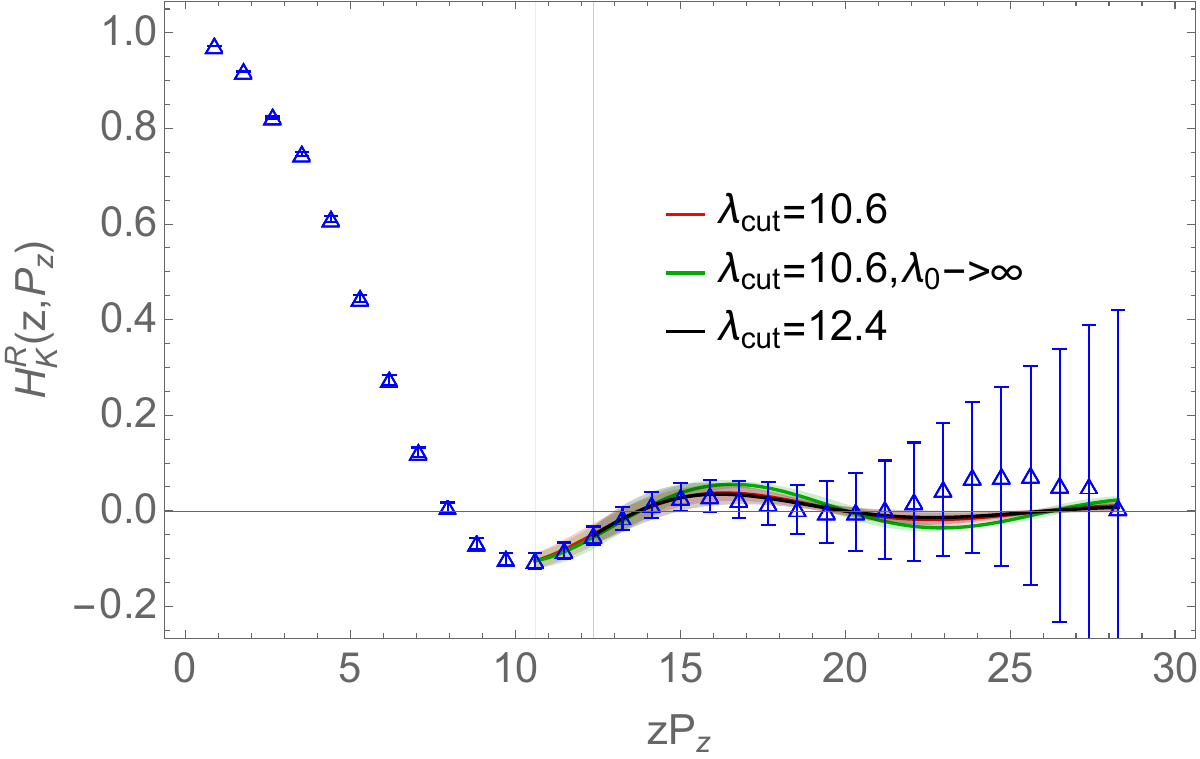}
    \includegraphics[width=0.45\textwidth]{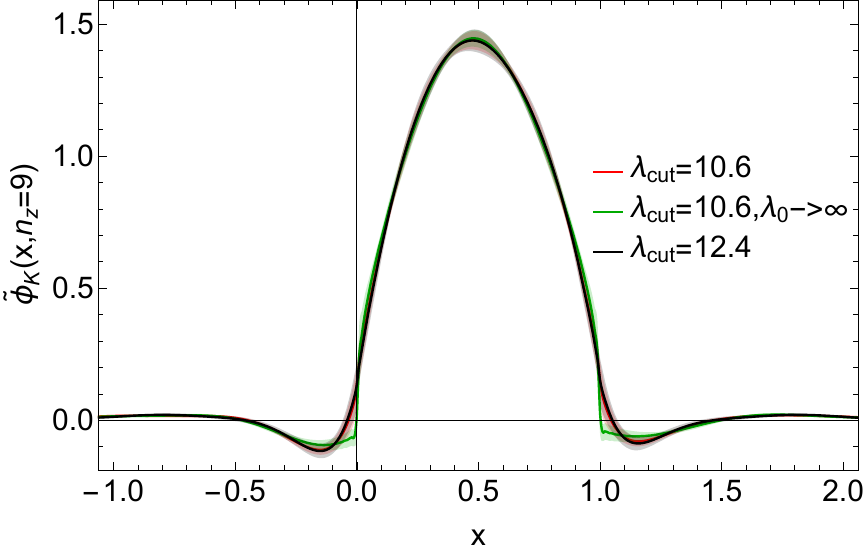}
    \caption{Long-tail extrapolation (left) and the corresponding $x$-dependent quasi-DA (right) for pion (top) and kaon (bottom). The different long-tail models introduce percent-level systematic uncertainties in the $x$ dependence. }
    \label{fig:longtail_ext}
\end{figure*}

The results are shown in Fig.~\ref{fig:longtail_ext}.  {The long-tail extrapolation will eventually introduce a small systematic uncertainties to the light-cone DA near $x=0.5$ in our case. Compared to the pion quasi-DA, the more precise long-range correlation in kaon quasi-DA helps reduce such a systematic uncertainty.}

\section{Matching to the light-cone DA}\label{sec:matching}
\subsection{Matching with small-momentum logarithm resummation}
We can then extract the light-cone DA from the quasi-DA through an inverse matching: 
\begin{align}
\label{eq:matching_mom}
 \phi(x,\mu)=&
\int_{-\infty}^{\infty} dy\ \mathcal{C}^{-1}(x,y,\mu,P_z)\tilde{\phi}(y,P_z) 
\nonumber\\&+ \mathcal{O}\left(\frac{\Lambda^2_\text{QCD}}{x^2P^2_z},\frac{\Lambda^2_\text{QCD}}{(1-x)^2P^2_z}\right),
\end{align}
where the power corrections is improved to quadratic order in $\Lambda^2_\text{QCD}/P^2_z$ with LRR, the perturbative matching kernel $\mathcal{C}(x,y,\mu,P_z)$ has been calculated at NLO~\cite{Liu:2018tox}.
Several improvements on the NLO matching kernel have been proposed to increase the accuracy of the perturbative matching in LaMET, including the LRR~\cite{Zhang:2023bxs,Holligan:2023rex} that eliminates the linear power correction, and the RGR~\cite{Su:2022fiu} and threshold resummation~\cite{Ji:2023pba} that resums the small-momentum logarithms. 

Our other recent work has derived the formalism for DA to resum all the small-momentum logarithms in the threshold limit $|x-y|\to 0$~\cite{Baker:2024zcd}, where the matching kernel can be factorized into the Sudakov factor $H(xP_z,\bar{x}P_z,\mu)$ and jet function $J(|x-y|P_z,\mu)$~\cite{Ji:2023pba},
\begin{align}\label{eq:tr_factorizaion}
    \mathcal{C}(x,y,\mu,P_z)\xrightarrow{x\to y}& H(xP_z,\bar{x}P_z,\mu)\otimes J(|x-y|P_z,\mu)\nonumber\\&+\mathcal{O}((y-x)^0).
\end{align}
In the threshold limit, the Sudakov factor 
\begin{align}
    H(xP_z,\bar{x}P_z,\mu)=\mathcal{F}[C^{-{\rm sgn}(z)}(xP_z,\mu)C^{{\rm sgn}(z)}(\bar{x}P_z,\mu)](x-y)
\end{align} 
incorporates the hard-collinear gluon modes connected to the external quark (antiquark) lines with momentum $xP_z$ ($\bar{x}P_z$), and the jet function $J(|x-y|P_z,\mu)=\mathcal{F}[\tilde{J}(z,\mu)](x-y)$ absorbs the soft gluon modes with momentum $|x-y|P_z$~\cite{Ji:2023pba,Baker:2024zcd}. They each follows individual RG equations
\begin{align}
    &\frac{\partial\ln C^{\pm}(p_z,\mu)}{\partial\ln\mu}=\frac{1}{2}\Gamma_{\rm cusp}(\alpha_s)\left(\ln\frac{4p_z^2}{\mu^2}\pm i\pi\right)+\gamma_c(\alpha_s),\nonumber\\
    &\frac{\partial\ln \tilde{J}(z,\mu)}{\partial\ln\mu}=\Gamma_{\rm cusp}(\alpha_s)\ln\frac{z^2\mu^2e^{2\gamma_E}}{4}-\gamma_J(\alpha_s),
\end{align}
where $\Gamma_{\rm cusp}, \gamma_c,\gamma_J$ are the corresponding anomalous dimensions~\cite{Su:2022fiu,Avkhadiev:2023poz,Baker:2024zcd}.

Solving these RG equations, we can obtain the resummed threshold components in the matching kernel.
\begin{align}
    JH_{\rm TR}&(\mu_i,\mu_{h_1},\mu_{h_2},\mu)=J_{\rm TR}(\mu_i,\mu)\otimes H_{\rm TR}(\mu_{h_1},\mu_{h_2},\mu),
\end{align}
where $\mu_i$ is the initial scale of jet function, and $\mu_{h_1}, \mu_{h_2}$ are the initial scale of the quark (antiquark) Sudakov factor, when solving the RG equations. 
Here the convolution order of $H$ and $J$ does not affect the threshold limit because they're multiplicative in coordinate space. So we average them and consider the difference between the two choices as a systematic error in our final results. The initial scales of the resummation are suggested to be~\cite{Baker:2024zcd},
\begin{align}
    \mu_i=2\min[x,\bar{x}]P_z,\quad 
    &\mu_{h_1}=2xP_z,\quad \mu_{h_2}=2\bar{x}P_z.
\end{align}

The full matching kernel is then resummed by first subtracting the threshold components $JH_{\rm NLO}$ at fixed order, then add back the resummed threshold terms $JH_{\rm TR}$,
\begin{align}%\label{eq:resummed_kernel}
    \mathcal{C}_{\rm TR}(\mu)=JH_{\rm TR}(\mu_i,\mu_{h_1},\mu_{h_2},\mu)\otimes JH^{-1}_{\rm NLO}(\mu)\otimes \mathcal{C}_{\rm NLO}(\mu),
\end{align}
and the resummed inverse matching kernel can be obtained as
\begin{align}%\label{eq:resummed_kernel}
    \mathcal{C}^{-1}_{\rm TR}(\mu)=\mathcal{C}^{-1}_{\rm NLO}(\mu)\otimes JH_{\rm NLO}(\mu)\otimes JH^{-1}_{\rm TR}(\mu_i,\mu_{h_1},\mu_{h_2},\mu).
\end{align}

\begin{figure}[!htbp]
    \centering
    \includegraphics[width=0.45\textwidth]{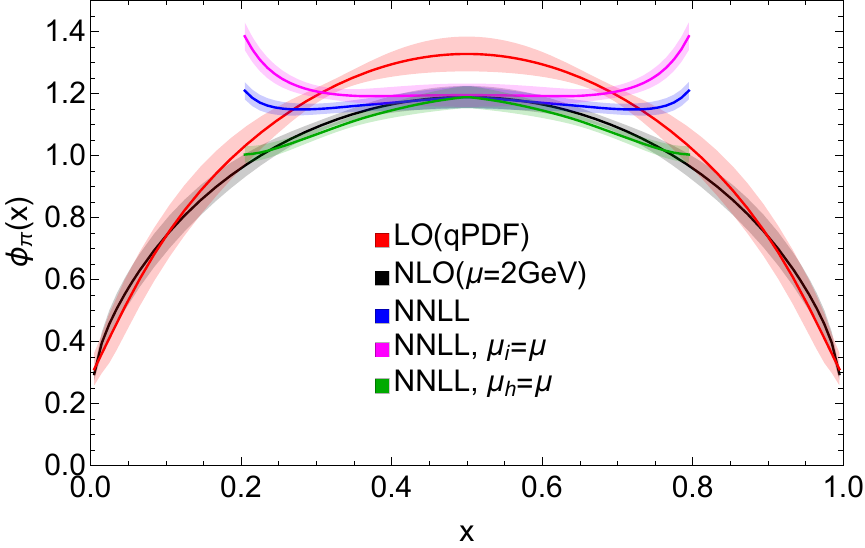}
    \includegraphics[width=0.45\textwidth]{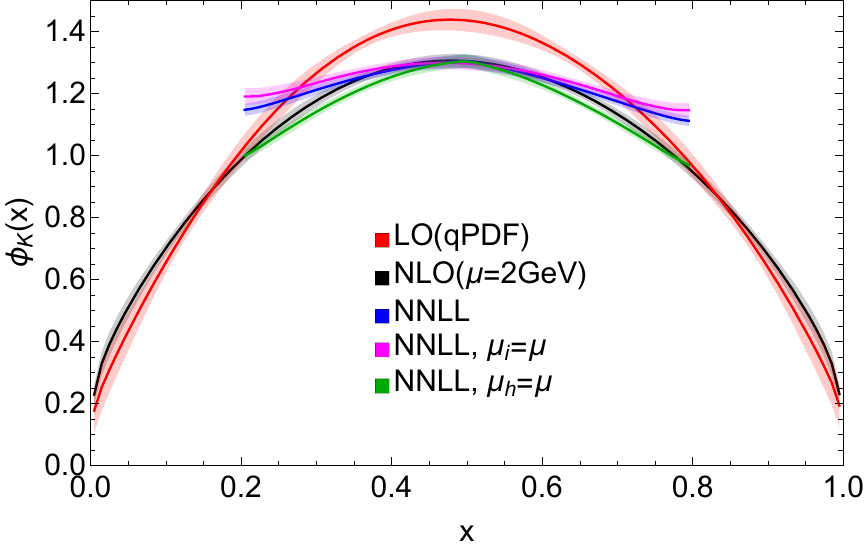}
    \caption{The fixed-order matching and the resummation effects on pion (top) and kaon (bottom) DA. The resummation effects of Sudakov factor and jet function turn out to be opposite.}
    \label{fig:matching_tr_effects}
\end{figure}

\begin{figure}[!htbp]
    \centering
    \includegraphics[width=0.45\textwidth]{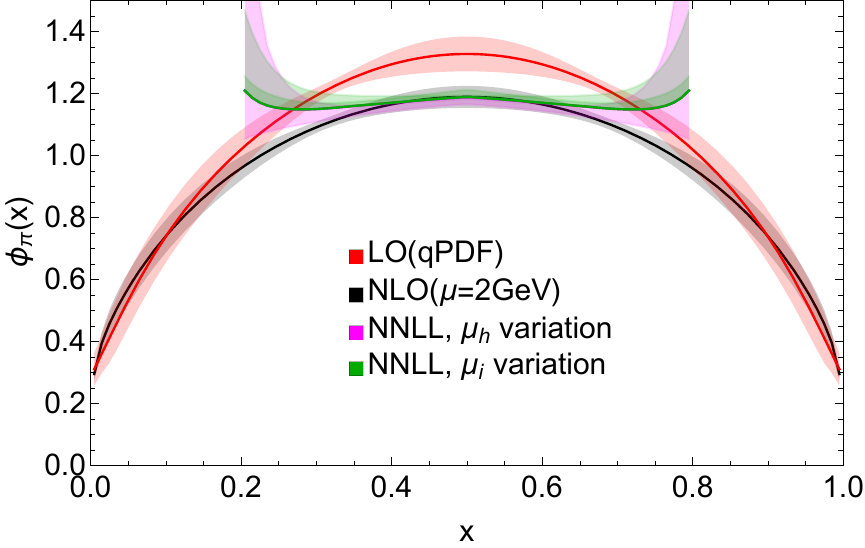}
    \includegraphics[width=0.45\textwidth]{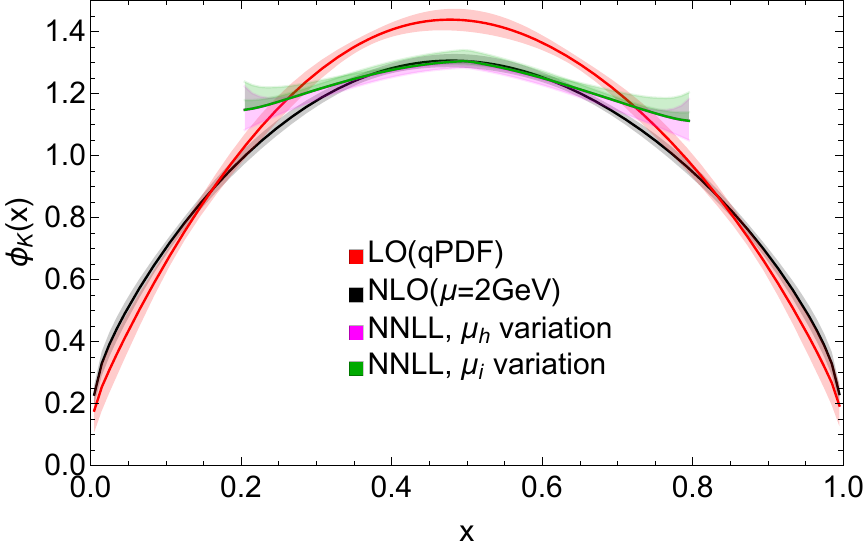}
    \caption{Scale variation for pion (top) and kaon (bottom). The uncertainty is small near $x=0.5$, but grows fast when approaching the endpoint regions. The kaon data are measured at a larger momentum, which allows us to access $x$ region closer to the endpoints. }
    \label{fig:scale_variation}
\end{figure}

Setting $\mu_i=\mu$ or $\mu_{h_1}=\mu_{h_2}=\mu$, we can examine the effects of resumming the Sudakov factors or the jet function separately, as shown in Fig.~\ref{fig:matching_tr_effects}. We find that the resummation of Sudakov factor and jet function have opposite effects on the matching.
To estimate the uncertainty from higher-order perturbation theory, we vary the initial scale of resummation by a factor of $\{\sqrt{2}^{-1},\sqrt{2}\}$. When the contribution from large logarithmic terms at higher orders of perturbation theory are significant, the resummed results will become sensitive to the choice of initial scales, indicating that the perturbation theory does not work or converge, so the matched DA is no longer reliable. We show the scale variation for the hard scale $\mu_h$ and semihard scale $\mu_i$ in Fig.~\ref{fig:scale_variation}. The scale dependence are smaller near $x=0.5$, but when approaching the endpoints, it becomes very large, indicating that the perturbation theory breaks down.  We find that the scale variation for kaon DA from $P_z=2.3$~GeV data is still controllable at $x\approx 0.2$. For pion with lower momentum $P_z=1.8$~GeV, the uncertainty grows much faster, thus the reliable range shrinks. Calculating the same observable with larger hadron momentum helps extend the range of reliability for our calculation. 

\begin{figure}[!htbp]
    \centering
    \includegraphics[width=0.45\textwidth]{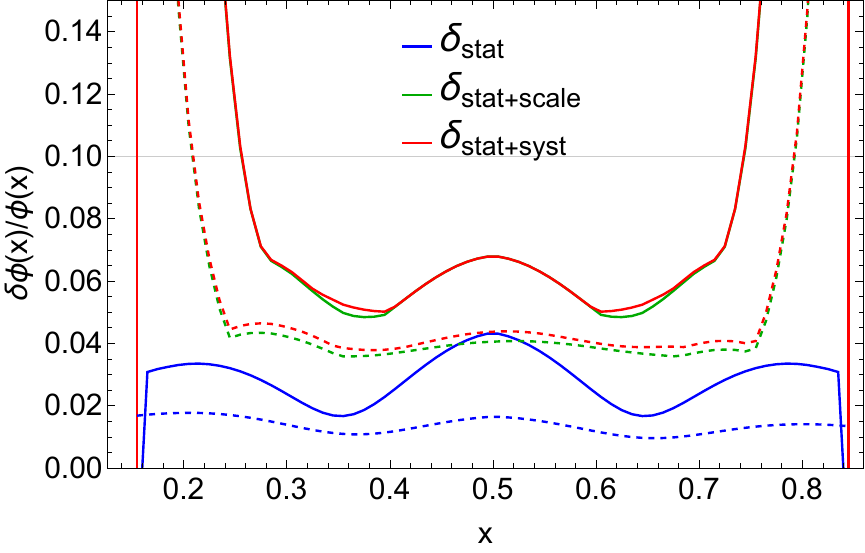}
    \caption{Relative uncertainties of pion (solid lines) and kaon (dashed lines) DAs as a function of $x$. }
    \label{fig:err_estimation}
\end{figure}

We consider the systematic error from the choice of $z_s\in[2a,3a]$ in the hybrid scheme, the model-dependence in the long-tail extrapolation, and the scale variation $c\in\{\sqrt{2}^{-1},1,\sqrt{2}\}$ in the hard and semihard scales with the replacement $\mu_{i,h}\to c\mu_{i,h}$. For each of these choices, we obtain $\phi_i(x)$ with a statistical error band. The systematic error band is obtained by requiring the total error band to cover all $\phi_i(x)$ results.
Figure~\ref{fig:err_estimation} shows the relative uncertainties as a function of $x$. Truncating at $10\%$ overall uncertainty, we find that the reliable range of our calculation is roughly $x\in[0.25,0.75]$ for pion, and $x\in[0.2,0.8]$ for kaon.

\begin{figure*}[!htbp]
    \centering
    \includegraphics[width=0.32\textwidth]{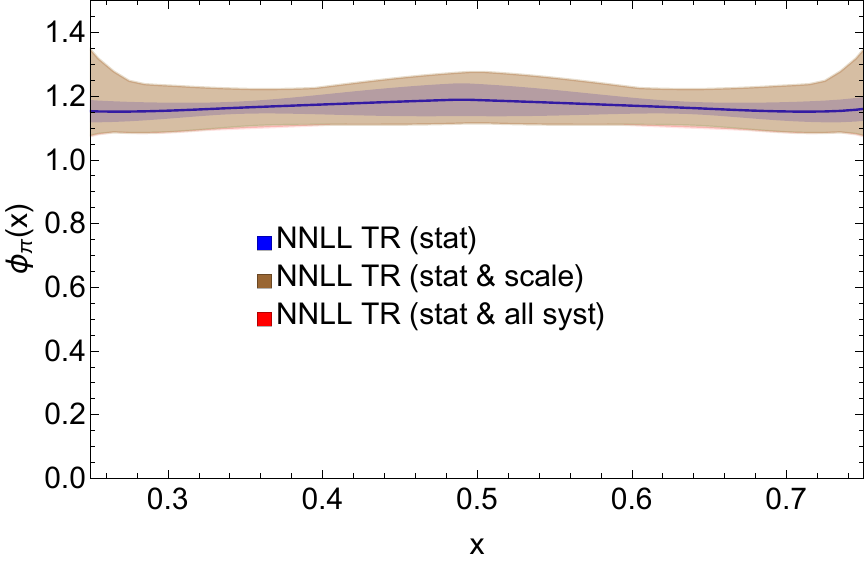}
    \includegraphics[width=0.32\textwidth]{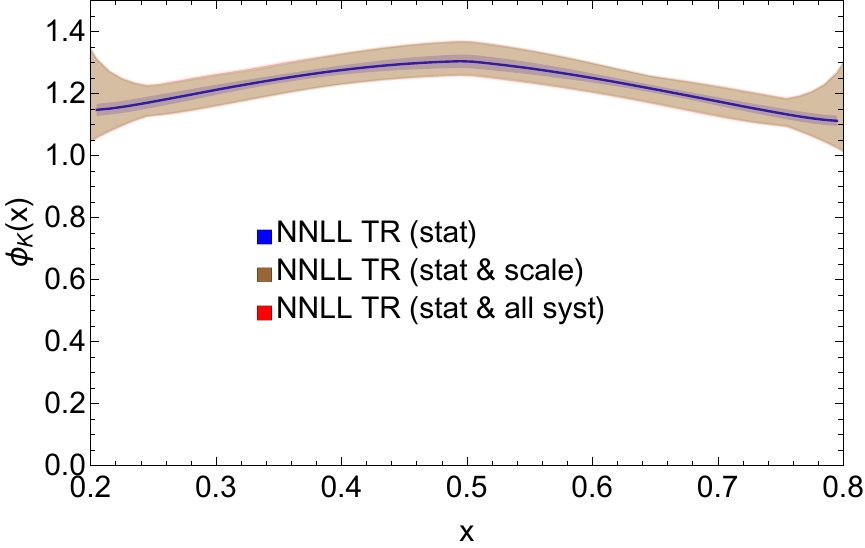}
    \includegraphics[width=0.32\textwidth]{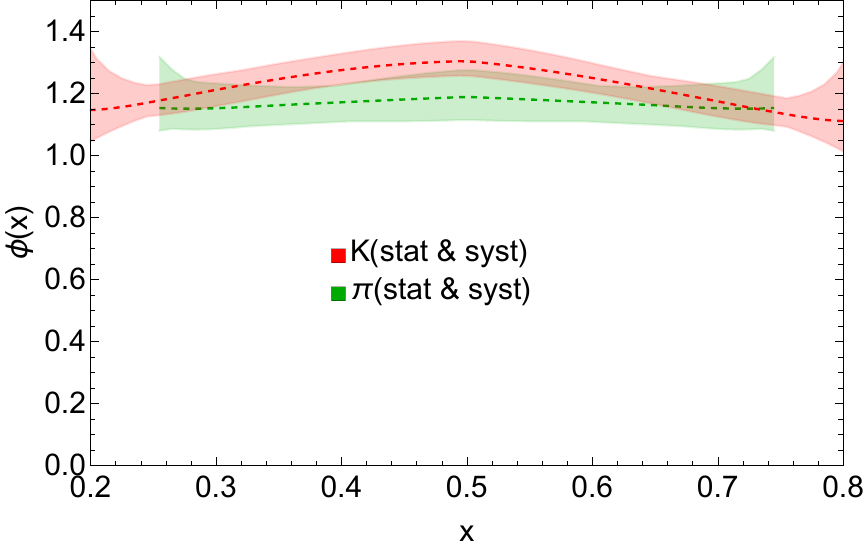}
    \caption{Final light-cone DA of pion (left) and kaon (mid), including scale variation and systematic errors. And a comparison between the two mesons are shown in the right panel. The kaon DA is narrower than the pion DA and slightly asymmetric.}
    \label{fig:final_result}
\end{figure*}

Combining these results, we show the final results of pion DA and kaon DA and their comparison in Fig.~\ref{fig:final_result}. Note that we only work on one single lattice spacing with mixed-action fermions. Thus the error band is not including any estimation of lattice artifacts, such as the discretization effects, finite volume effects, and mixed-action effects. According to previous lattice calculations, the discretization effects are indeed important~\cite{Zhang:2020gaj,LatticeParton:2022zqc}, and the finite volume effect could be small~\cite{Lin:2019ocg} since the boosted hadrons have relatively short wavelength. These effects also need to be carefully examined, in combination with our theoretical improvements, to present a reliable lattice benchmark for the meson DAs.

\subsection{Higher moments from complementarity}\label{app:complementarity}
With the LaMET method, we have calculated the pion DA for a range of $x\in[0.25,0.75]$ for pion and $x\in[0.2,0.8]$ for kaon. Beyond this range, the perturbation theory no longer converges, and the power corrections in Eq.~\eqref{eq:matching_mom} are increasingly important. Both effects prevent us from extracting the distribution closer to the endpoint with current hadron momentum $P_z\sim 2$~GeV. Although the $x$-dependence is not directly calculable in this region, it has been proposed to constrain the endpoint regions with the global information of DA, such as the lower moments or short distance correlations, extracted from SDF~\cite{Ji:2022ezo,Holligan:2023rex}. In coordinate space, the perturbation theory still works at $z\ll\Lambda_{\rm QCD}^{-1}$, and the power corrections $\mathcal{O}(z^n\Lambda^n_{\rm QCD})$ are also suppressed at short range. Thus the short distance correlations, along with the already determined mid-$x$ distribution, can be utilized to better constrain the distribution in the endpoint regions, thus allowing us to roughly estimate the higher moments of DA. 

Here we use a simple power-law model for the distribution near the endpoint, $\phi(x\to0,1)\propto A x^m(1-x)^n$. Then we get a full $\phi(x,\mu,m,n)$ in $[0,1]$ depending on the unknown parameters $m$ and $n$. {For the pion we set $m=n$ since its DA is symmetric around $x=0.5$.} By Fourier transforming $\phi(x,\mu,m,n)$ to coordinate space, and implementing the RG-improved matching using SDF, we obtain the quasi-DA correlations $H(z,m,n)$. 
%\YZ{Has RGR and TR been used in the following formula? We probably should clarify.}
\begin{align}
    H(z,m,n)&=\int d\nu \mathcal{Z}(\nu,z^2,\mu,zP_z) \times \nonumber\\
    &\int P_zdz\ e^{i(\frac{1}{2}-x)\nu zP_z}\phi(x,\mu,m,n),
\end{align}
where $\mathcal{Z}(\nu,z^2,\mu,zP_z)$ is the matching kernel for coordinate space correlations~\cite{Radyushkin:2019owq,Gao:2022vyh,Holligan:2023rex}.
Then the parameters $m,n$ can be obtained by fitting the lattice matrix elements $H^R(z)$ to $H(z,m,n)$. The parameters of pion DA are fitted using the real part of the lattice matrix elements because we have enforced the isospin symmetry, while the kaon DA model are constrained by both real and imaginary parts of the lattice matrix elements.

After determining the parameters $m$ and $n$, we can estimate the higher moments of DA. The Mellin moments $\langle\xi^i\rangle$ and Gegenbauer moments $a_i$ are calculated by
\begin{align}
    &\langle\xi^i\rangle=\int_0^1 dx\phi(x)(2x-1)^i,\\
    &a_i=\int_0^1 dx\phi(x) C^{3/2}_i[2x-1]\frac{4(i+3/2)}{3(i+1)(i+2)},
\end{align}
where $C^{3/2}_i[2x-1]$ are the Gegenbauer polynomials.
The results are summarized in Table~\ref{tab:moments}. 
\begin{table*}[]
    \centering
    \vline\begin{tabular}{c|c|c|c|c|c|c|c|c|c}
    \hline
             & $m$ & $n$ & $a_2$ & $a_4$ & $a_6$ &$\langle \xi\rangle$ &$\langle \xi^2\rangle$ & $\langle \xi^4\rangle$ & $\langle \xi^6\rangle$ \\
             \hline
         % $K$& 0.59(7)& 0.58(7)&  0.134(22) & 0.065(12) & 0.035(7) &0.245(8) & 0.123(6) & 0.077(5)  \\
         $K$& 0.62(7)& 0.58(7)&  0.114(20) & 0.037(11) & 0.019(5) &0.001(10)&0.237(7) & 0.115(6) & 0.070(5)  \\
         \hline
         $\pi$& 0.31(6)& 0.31(6) & 0.196(32) & 0.085(26) & 0.056(15) &0& 0.267(11) & 0.139(10) & 0.090(8) \\
         \hline
    \end{tabular}\vline
    \caption{Mellin moments, $\langle\xi^i\rangle$, and Gegenbauer moments, $a_i$, estimated from modeling the endpoint region of pion and kaon DA  with $\phi(x_0<x<1-x_0)$ determined by LaMET calculation.}
    \label{tab:moments}
\end{table*}
The second moment of our kaon DA is consistent with the local twist-2 operators by  RQCD collaboration~\cite{RQCD:2019osh} at the physical limit, but the second moment of our pion DA is much larger than their calculation. Our moment extraction highly depends on the short-distance correlations, which are more sensitive to discretization effects, thus the errors presented in the table are missing an important source of lattice artifacts. The first moment of out kaon DA is almost vanishing, very different from the RQCD calculation. It could come from the kinetic contamination of the imaginary part in Eq.~\eqref{eq:LRZdc} not being fully subtracted in our calculation due to the remaining $P_z$-dependence, making our estimation of the kaon skewness less reliable. If we first ignore the lattice artifacts and just compare different theoretical approaches,
in a previous work on the moment analysis of the same pion DA data~\cite{Gao:2022vyh} with NLO matching,
%\YZ{with only NLO matching?}, 
it is suggested that the power-law parameter $m$ lies between $0.2$ and $0.32$ with different analysis choices, consistent with our estimate $m=0.31(6)$ in this calculation. Our second moment and fourth moment are also in 2$\sigma$ agreement with their analysis of $\langle\xi^2\rangle=0.287(6)(6)$ and $\langle\xi^4\rangle=0.14(3)(3)$. 
% \textcolor{red}{SM: Need comparisons of pion DA $m$, $n$ and different moments with \cite{Gao:2022vyh}.}
Note that the fitted second Mellin moments $\langle \xi^2\rangle$ for the pion and kaon data in this approach agree with those fitted from RG-invariant ratios in Fig.~\ref{fig:moment_fit}. Such consistency in the lowest moments provide us more confidence about our estimate of higher moments on this ensemble. From both moments and the $x$-dependence, the pion DA is flatter than kaon, showing the non-negligible quark mass dependence of the meson DAs. The kaon is only slightly skewed as $m$ and $n$ are very close, and the first Mellin moment is consistent with zero. 

\begin{figure}
\centering
    \includegraphics[width=0.5\textwidth]{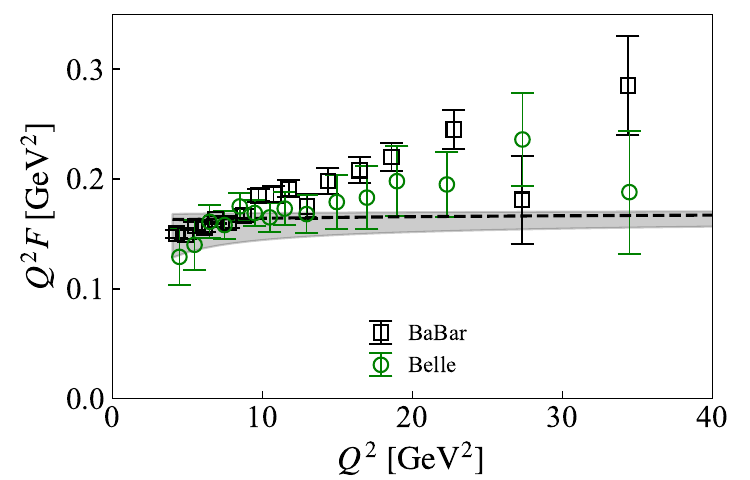}
	\caption{The pion transition form factor predicted from collinear factorization. The band comes from the scale variation of $\mu\in[Q/2,2Q]$, while the dashed line represent the result from $\mu=Q$.\label{fig:TransitionFF}}
\end{figure}

Higher moments estimated from this approach can be used as inputs to the pion and kaon phenomenology. An example is to calculate the pion transition form factor $F_{\pi\gamma\gamma^*}(Q^2)$ at large $Q^2$, which can be factorized as,
\begin{align}
    F(Q^2)=\frac{\sqrt{2}f_\pi}{6Q^2}\int_0^1\phi(x,\mu)T(x,\mu,Q^2),
\end{align}
where the hard kernel $T(x,\mu,Q^2)$ has been calculated up to 2-loop order ~\cite{Braun:2021grd}. The convolution has also been formulated in terms of Gagenbauer coefficients with the hard coefficients $c_n$ provided up to 2-loop~\cite{Braun:2021grd},
\begin{align}
    Q^2F(Q^2)=\sqrt{2}f_\pi\sum_{n=0,2,\dots}a_n(\mu) c_n(\mu,Q).
\end{align}
We truncate the above equation at $n=6$ and show the result from our estimate of moments in Fig.~\ref{fig:TransitionFF}. The scale variation is examined by choosing $\mu\in\{Q/2,Q,2Q\}$, with the corresponding $a_n(\mu)$ evolved from $2$~GeV to $\mu$ using their anomalous dimensions up to 3 loops~\cite{Braun:2017cih}. Our band is in good agreement with data from Belle~\cite{Belle-II:2018jsg} from $Q^2>4{\rm ~GeV}^2$. 

Note that our estimation of higher moments are model-dependent. For our pion momentum $P_z\approx 1.8$~GeV, LaMET only provides reliable prediction in a range of $x\in[0.25,0.75]$. In principle, this range is not reaching close enough to the endpoint for us to model the DA as a simple power-law function. Moreover, since the lattice spacing is not fine enough, we only have three data points of spatial correlation function at $z\leq 3a$ that stays within the perturbative region, thus a complicated model could hardly be determined by this approach. To provide a more reliable estimate for the overall shape and higher moments of DA, we should include more calculations on finer lattice spacings and with higher hadron momenta.

\section{Conclusion}~\label{sec:conlusion}

In this work, we present a lattice QCD calculation of the $x$-dependent pion and kaon DAs in the framework of LaMET. This calculation is performed on a fine lattice of $a=0.076$~fm at physical pion mass, and has used the state-of-the-art analysis methods, including the recently developed resummation methods of small-momentum logarithms. With the pion boosted to $1.8$~GeV and kaon boosted to $2.3$~GeV, we are able to calculate a range of $x\in[x_0,1-x_0]$ with $x_0=0.25$ for pion and $x_0=0.2$ for kaon with theoretical systematic errors under control. Beyond this range, the perturbation theory and the power-expansion of the factorization theorem are no longer reliable, thus the DA cannot be directly accessed. Our results suggest the pion DA is flatter than the kaon DA, and the asymmetry in the kaon DA is small. Using complementarity, we estimate higher moments of the pion and kaon DAs by combining our calculation with short-distance factorization. The second moments are consistent with the OPE fits to the RG-invariant ratios of the matrix elements.

\section*{Acknowledgment}

This material is based upon work supported by The U.S. Department of Energy, Office of Science, Office of Nuclear Physics through Contract No.~DE-SC0012704, Contract No.~DE-AC02-06CH11357, and within the frameworks of Scientific Discovery through Advanced Computing (SciDAC) award Fundamental Nuclear Physics at the Exascale and Beyond and the Topical Collaboration in Nuclear Theory 3D quark-gluon structure of hadrons: mass, spin, and tomography. YZ was partially supported by the 2023 Physical Sciences and Engineering (PSE) Early Investigator Named Award program at Argonne National Laboratory.
%This work was supported in part by the U.S. Department of Energy, Office of Science, Office of Workforce Development for Teachers and Scientists (WDTS) under the Science Undergraduate Laboratory Internships Program (SULI). 

This research used awards of computer time provided by: The INCITE program at Argonne Leadership Computing Facility, a DOE Office of Science User Facility operated under Contract No.~DE-AC02-06CH11357; the ALCC program at the Oak Ridge Leadership Computing Facility, which is a DOE Office of Science User Facility supported under Contract DE-AC05-00OR22725; the National Energy Research Scientific Computing Center, a DOE Office of Science User Facility
supported by the Office of Science of the U.S. Department of Energy under Contract No. DE-AC02-05CH11231 using NERSC award NP-ERCAP0028137. Computations for this work were carried out in part on facilities of the USQCD Collaboration, which are funded by the Office of Science of the U.S. Department of Energy. Part of the data analysis are carried out on Swing, a high-performance computing cluster operated by the Laboratory Computing Resource Center at Argonne National Laboratory.


\begin{thebibliography}{83}
\expandafter\ifx\csname natexlab\endcsname\relax\def\natexlab#1{#1}\fi
\expandafter\ifx\csname bibnamefont\endcsname\relax
  \def\bibnamefont#1{#1}\fi
\expandafter\ifx\csname bibfnamefont\endcsname\relax
  \def\bibfnamefont#1{#1}\fi
\expandafter\ifx\csname citenamefont\endcsname\relax
  \def\citenamefont#1{#1}\fi
\expandafter\ifx\csname url\endcsname\relax
  \def\url#1{\texttt{#1}}\fi
\expandafter\ifx\csname urlprefix\endcsname\relax\def\urlprefix{URL }\fi
\providecommand{\bibinfo}[2]{#2}
\providecommand{\eprint}[2][]{\url{#2}}

\bibitem[{\citenamefont{Beneke et~al.}(1999)\citenamefont{Beneke, Buchalla,
  Neubert, and Sachrajda}}]{Beneke:1999br}
\bibinfo{author}{\bibfnamefont{M.}~\bibnamefont{Beneke}},
  \bibinfo{author}{\bibfnamefont{G.}~\bibnamefont{Buchalla}},
  \bibinfo{author}{\bibfnamefont{M.}~\bibnamefont{Neubert}}, \bibnamefont{and}
  \bibinfo{author}{\bibfnamefont{C.~T.} \bibnamefont{Sachrajda}},
  \bibinfo{journal}{Phys. Rev. Lett.} \textbf{\bibinfo{volume}{83}},
  \bibinfo{pages}{1914} (\bibinfo{year}{1999}), \eprint{hep-ph/9905312}.

\bibitem[{\citenamefont{Beneke et~al.}(2001)\citenamefont{Beneke, Buchalla,
  Neubert, and Sachrajda}}]{Beneke:2001ev}
\bibinfo{author}{\bibfnamefont{M.}~\bibnamefont{Beneke}},
  \bibinfo{author}{\bibfnamefont{G.}~\bibnamefont{Buchalla}},
  \bibinfo{author}{\bibfnamefont{M.}~\bibnamefont{Neubert}}, \bibnamefont{and}
  \bibinfo{author}{\bibfnamefont{C.~T.} \bibnamefont{Sachrajda}},
  \bibinfo{journal}{Nucl. Phys. B} \textbf{\bibinfo{volume}{606}},
  \bibinfo{pages}{245} (\bibinfo{year}{2001}), \eprint{hep-ph/0104110}.

\bibitem[{\citenamefont{Collins et~al.}(1997)\citenamefont{Collins, Frankfurt,
  and Strikman}}]{Collins:1996fb}
\bibinfo{author}{\bibfnamefont{J.~C.} \bibnamefont{Collins}},
  \bibinfo{author}{\bibfnamefont{L.}~\bibnamefont{Frankfurt}},
  \bibnamefont{and} \bibinfo{author}{\bibfnamefont{M.}~\bibnamefont{Strikman}},
  \bibinfo{journal}{Phys. Rev. D} \textbf{\bibinfo{volume}{56}},
  \bibinfo{pages}{2982} (\bibinfo{year}{1997}), \eprint{hep-ph/9611433}.

\bibitem[{\citenamefont{Behrend et~al.}(1991)}]{CELLO:1990klc}
\bibinfo{author}{\bibfnamefont{H.~J.} \bibnamefont{Behrend}}
  \bibnamefont{et~al.} (\bibinfo{collaboration}{CELLO}), \bibinfo{journal}{Z.
  Phys. C} \textbf{\bibinfo{volume}{49}}, \bibinfo{pages}{401}
  (\bibinfo{year}{1991}).

\bibitem[{\citenamefont{Stewart}(2003)}]{Stewart:2003gt}
\bibinfo{author}{\bibfnamefont{I.~W.} \bibnamefont{Stewart}}, in
  \emph{\bibinfo{booktitle}{{38th Rencontres de Moriond on QCD and High-Energy
  Hadronic Interactions}}} (\bibinfo{year}{2003}), \eprint{hep-ph/0308185}.

\bibitem[{\citenamefont{Li and Mishima}(2011)}]{Li:2009wba}
\bibinfo{author}{\bibfnamefont{H.-n.} \bibnamefont{Li}} \bibnamefont{and}
  \bibinfo{author}{\bibfnamefont{S.}~\bibnamefont{Mishima}},
  \bibinfo{journal}{Phys. Rev. D} \textbf{\bibinfo{volume}{83}},
  \bibinfo{pages}{034023} (\bibinfo{year}{2011}), \eprint{0901.1272}.

\bibitem[{\citenamefont{Li et~al.}(2012)\citenamefont{Li, Lu, and
  Yu}}]{Li:2012cfa}
\bibinfo{author}{\bibfnamefont{H.-n.} \bibnamefont{Li}},
  \bibinfo{author}{\bibfnamefont{C.-D.} \bibnamefont{Lu}}, \bibnamefont{and}
  \bibinfo{author}{\bibfnamefont{F.-S.} \bibnamefont{Yu}},
  \bibinfo{journal}{Phys. Rev. D} \textbf{\bibinfo{volume}{86}},
  \bibinfo{pages}{036012} (\bibinfo{year}{2012}), \eprint{1203.3120}.

\bibitem[{\citenamefont{Gronberg et~al.}(1998)}]{CLEO:1997fho}
\bibinfo{author}{\bibfnamefont{J.}~\bibnamefont{Gronberg}} \bibnamefont{et~al.}
  (\bibinfo{collaboration}{CLEO}), \bibinfo{journal}{Phys. Rev. D}
  \textbf{\bibinfo{volume}{57}}, \bibinfo{pages}{33} (\bibinfo{year}{1998}),
  \eprint{hep-ex/9707031}.

\bibitem[{\citenamefont{Aubert et~al.}(2009)}]{BaBar:2009rrj}
\bibinfo{author}{\bibfnamefont{B.}~\bibnamefont{Aubert}} \bibnamefont{et~al.}
  (\bibinfo{collaboration}{BaBar}), \bibinfo{journal}{Phys. Rev. D}
  \textbf{\bibinfo{volume}{80}}, \bibinfo{pages}{052002}
  (\bibinfo{year}{2009}), \eprint{0905.4778}.

\bibitem[{\citenamefont{Altmannshofer et~al.}(2019)}]{Belle-II:2018jsg}
\bibinfo{author}{\bibfnamefont{W.}~\bibnamefont{Altmannshofer}}
  \bibnamefont{et~al.} (\bibinfo{collaboration}{Belle-II}),
  \bibinfo{journal}{PTEP} \textbf{\bibinfo{volume}{2019}},
  \bibinfo{pages}{123C01} (\bibinfo{year}{2019}), \bibinfo{note}{[Erratum: PTEP
  2020, 029201 (2020)]}, \eprint{1808.10567}.

\bibitem[{\citenamefont{Chang et~al.}(2013)\citenamefont{Chang, Cloet,
  Cobos-Martinez, Roberts, Schmidt, and Tandy}}]{Chang:2013pq}
\bibinfo{author}{\bibfnamefont{L.}~\bibnamefont{Chang}},
  \bibinfo{author}{\bibfnamefont{I.~C.} \bibnamefont{Cloet}},
  \bibinfo{author}{\bibfnamefont{J.~J.} \bibnamefont{Cobos-Martinez}},
  \bibinfo{author}{\bibfnamefont{C.~D.} \bibnamefont{Roberts}},
  \bibinfo{author}{\bibfnamefont{S.~M.} \bibnamefont{Schmidt}},
  \bibnamefont{and} \bibinfo{author}{\bibfnamefont{P.~C.} \bibnamefont{Tandy}},
  \bibinfo{journal}{Phys. Rev. Lett.} \textbf{\bibinfo{volume}{110}},
  \bibinfo{pages}{132001} (\bibinfo{year}{2013}), \eprint{1301.0324}.

\bibitem[{\citenamefont{Shi et~al.}(2014)\citenamefont{Shi, Chang, Roberts,
  Schmidt, Tandy, and Zong}}]{Shi:2014uwa}
\bibinfo{author}{\bibfnamefont{C.}~\bibnamefont{Shi}},
  \bibinfo{author}{\bibfnamefont{L.}~\bibnamefont{Chang}},
  \bibinfo{author}{\bibfnamefont{C.~D.} \bibnamefont{Roberts}},
  \bibinfo{author}{\bibfnamefont{S.~M.} \bibnamefont{Schmidt}},
  \bibinfo{author}{\bibfnamefont{P.~C.} \bibnamefont{Tandy}}, \bibnamefont{and}
  \bibinfo{author}{\bibfnamefont{H.-S.} \bibnamefont{Zong}},
  \bibinfo{journal}{Phys. Lett. B} \textbf{\bibinfo{volume}{738}},
  \bibinfo{pages}{512} (\bibinfo{year}{2014}), \eprint{1406.3353}.

\bibitem[{\citenamefont{de~Melo et~al.}(2016)\citenamefont{de~Melo, Ahmed, and
  Tsushima}}]{deMelo:2015yxk}
\bibinfo{author}{\bibfnamefont{J.~P. B.~C.} \bibnamefont{de~Melo}},
  \bibinfo{author}{\bibfnamefont{I.}~\bibnamefont{Ahmed}}, \bibnamefont{and}
  \bibinfo{author}{\bibfnamefont{K.}~\bibnamefont{Tsushima}},
  \bibinfo{journal}{AIP Conf. Proc.} \textbf{\bibinfo{volume}{1735}},
  \bibinfo{pages}{080012} (\bibinfo{year}{2016}), \eprint{1512.07260}.

\bibitem[{\citenamefont{Kronfeld and Photiadis}(1985)}]{Kronfeld:1984zv}
\bibinfo{author}{\bibfnamefont{A.~S.} \bibnamefont{Kronfeld}} \bibnamefont{and}
  \bibinfo{author}{\bibfnamefont{D.~M.} \bibnamefont{Photiadis}},
  \bibinfo{journal}{Phys. Rev. D} \textbf{\bibinfo{volume}{31}},
  \bibinfo{pages}{2939} (\bibinfo{year}{1985}).

\bibitem[{\citenamefont{Del~Debbio et~al.}(2003)\citenamefont{Del~Debbio,
  Di~Pierro, and Dougall}}]{DelDebbio:2002mq}
\bibinfo{author}{\bibfnamefont{L.}~\bibnamefont{Del~Debbio}},
  \bibinfo{author}{\bibfnamefont{M.}~\bibnamefont{Di~Pierro}},
  \bibnamefont{and} \bibinfo{author}{\bibfnamefont{A.}~\bibnamefont{Dougall}},
  \bibinfo{journal}{Nucl. Phys. B Proc. Suppl.} \textbf{\bibinfo{volume}{119}},
  \bibinfo{pages}{416} (\bibinfo{year}{2003}), \eprint{hep-lat/0211037}.

\bibitem[{\citenamefont{Braun et~al.}(2006)}]{Braun:2006dg}
\bibinfo{author}{\bibfnamefont{V.~M.} \bibnamefont{Braun}}
  \bibnamefont{et~al.}, \bibinfo{journal}{Phys. Rev. D}
  \textbf{\bibinfo{volume}{74}}, \bibinfo{pages}{074501}
  (\bibinfo{year}{2006}), \eprint{hep-lat/0606012}.

\bibitem[{\citenamefont{Arthur et~al.}(2011)\citenamefont{Arthur, Boyle,
  Brommel, Donnellan, Flynn, Juttner, Rae, and Sachrajda}}]{Arthur:2010xf}
\bibinfo{author}{\bibfnamefont{R.}~\bibnamefont{Arthur}},
  \bibinfo{author}{\bibfnamefont{P.~A.} \bibnamefont{Boyle}},
  \bibinfo{author}{\bibfnamefont{D.}~\bibnamefont{Brommel}},
  \bibinfo{author}{\bibfnamefont{M.~A.} \bibnamefont{Donnellan}},
  \bibinfo{author}{\bibfnamefont{J.~M.} \bibnamefont{Flynn}},
  \bibinfo{author}{\bibfnamefont{A.}~\bibnamefont{Juttner}},
  \bibinfo{author}{\bibfnamefont{T.~D.} \bibnamefont{Rae}}, \bibnamefont{and}
  \bibinfo{author}{\bibfnamefont{C.~T.~C.} \bibnamefont{Sachrajda}},
  \bibinfo{journal}{Phys. Rev. D} \textbf{\bibinfo{volume}{83}},
  \bibinfo{pages}{074505} (\bibinfo{year}{2011}), \eprint{1011.5906}.

\bibitem[{\citenamefont{Bali et~al.}(2017)\citenamefont{Bali, Braun,
  G\"ockeler, Gruber, Hutzler, Korcyl, Lang, and Sch\"afer}}]{Bali:2017ude}
\bibinfo{author}{\bibfnamefont{G.~S.} \bibnamefont{Bali}},
  \bibinfo{author}{\bibfnamefont{V.~M.} \bibnamefont{Braun}},
  \bibinfo{author}{\bibfnamefont{M.}~\bibnamefont{G\"ockeler}},
  \bibinfo{author}{\bibfnamefont{M.}~\bibnamefont{Gruber}},
  \bibinfo{author}{\bibfnamefont{F.}~\bibnamefont{Hutzler}},
  \bibinfo{author}{\bibfnamefont{P.}~\bibnamefont{Korcyl}},
  \bibinfo{author}{\bibfnamefont{B.}~\bibnamefont{Lang}}, \bibnamefont{and}
  \bibinfo{author}{\bibfnamefont{A.}~\bibnamefont{Sch\"afer}}
  (\bibinfo{collaboration}{RQCD}), \bibinfo{journal}{Phys. Lett. B}
  \textbf{\bibinfo{volume}{774}}, \bibinfo{pages}{91} (\bibinfo{year}{2017}),
  \eprint{1705.10236}.

\bibitem[{\citenamefont{Bali et~al.}(2019)\citenamefont{Bali, Braun, B\"urger,
  G\"ockeler, Gruber, Hutzler, Korcyl, Sch\"afer, Sternbeck, and
  Wein}}]{RQCD:2019osh}
\bibinfo{author}{\bibfnamefont{G.~S.} \bibnamefont{Bali}},
  \bibinfo{author}{\bibfnamefont{V.~M.} \bibnamefont{Braun}},
  \bibinfo{author}{\bibfnamefont{S.}~\bibnamefont{B\"urger}},
  \bibinfo{author}{\bibfnamefont{M.}~\bibnamefont{G\"ockeler}},
  \bibinfo{author}{\bibfnamefont{M.}~\bibnamefont{Gruber}},
  \bibinfo{author}{\bibfnamefont{F.}~\bibnamefont{Hutzler}},
  \bibinfo{author}{\bibfnamefont{P.}~\bibnamefont{Korcyl}},
  \bibinfo{author}{\bibfnamefont{A.}~\bibnamefont{Sch\"afer}},
  \bibinfo{author}{\bibfnamefont{A.}~\bibnamefont{Sternbeck}},
  \bibnamefont{and} \bibinfo{author}{\bibfnamefont{P.}~\bibnamefont{Wein}}
  (\bibinfo{collaboration}{RQCD}), \bibinfo{journal}{JHEP}
  \textbf{\bibinfo{volume}{08}}, \bibinfo{pages}{065} (\bibinfo{year}{2019}),
  \bibinfo{note}{[Addendum: JHEP 11, 037 (2020)]}, \eprint{1903.08038}.

\bibitem[{\citenamefont{Braun and M\"uller}(2008)}]{Braun:2007wv}
\bibinfo{author}{\bibfnamefont{V.}~\bibnamefont{Braun}} \bibnamefont{and}
  \bibinfo{author}{\bibfnamefont{D.}~\bibnamefont{M\"uller}},
  \bibinfo{journal}{Eur. Phys. J. C} \textbf{\bibinfo{volume}{55}},
  \bibinfo{pages}{349} (\bibinfo{year}{2008}), \eprint{0709.1348}.

\bibitem[{\citenamefont{Braun et~al.}(2015)\citenamefont{Braun, Collins,
  G\"ockeler, P\'erez-Rubio, Sch\"afer, Schiel, and Sternbeck}}]{Braun:2015axa}
\bibinfo{author}{\bibfnamefont{V.~M.} \bibnamefont{Braun}},
  \bibinfo{author}{\bibfnamefont{S.}~\bibnamefont{Collins}},
  \bibinfo{author}{\bibfnamefont{M.}~\bibnamefont{G\"ockeler}},
  \bibinfo{author}{\bibfnamefont{P.}~\bibnamefont{P\'erez-Rubio}},
  \bibinfo{author}{\bibfnamefont{A.}~\bibnamefont{Sch\"afer}},
  \bibinfo{author}{\bibfnamefont{R.~W.} \bibnamefont{Schiel}},
  \bibnamefont{and}
  \bibinfo{author}{\bibfnamefont{A.}~\bibnamefont{Sternbeck}},
  \bibinfo{journal}{Phys. Rev. D} \textbf{\bibinfo{volume}{92}},
  \bibinfo{pages}{014504} (\bibinfo{year}{2015}), \eprint{1503.03656}.

\bibitem[{\citenamefont{Bali et~al.}(2018{\natexlab{a}})\citenamefont{Bali,
  Braun, Gl\"a\ss{}le, G\"ockeler, Gruber, Hutzler, Korcyl, Sch\"afer, Wein,
  and Zhang}}]{Bali:2018spj}
\bibinfo{author}{\bibfnamefont{G.~S.} \bibnamefont{Bali}},
  \bibinfo{author}{\bibfnamefont{V.~M.} \bibnamefont{Braun}},
  \bibinfo{author}{\bibfnamefont{B.}~\bibnamefont{Gl\"a\ss{}le}},
  \bibinfo{author}{\bibfnamefont{M.}~\bibnamefont{G\"ockeler}},
  \bibinfo{author}{\bibfnamefont{M.}~\bibnamefont{Gruber}},
  \bibinfo{author}{\bibfnamefont{F.}~\bibnamefont{Hutzler}},
  \bibinfo{author}{\bibfnamefont{P.}~\bibnamefont{Korcyl}},
  \bibinfo{author}{\bibfnamefont{A.}~\bibnamefont{Sch\"afer}},
  \bibinfo{author}{\bibfnamefont{P.}~\bibnamefont{Wein}}, \bibnamefont{and}
  \bibinfo{author}{\bibfnamefont{J.-H.} \bibnamefont{Zhang}},
  \bibinfo{journal}{Phys. Rev. D} \textbf{\bibinfo{volume}{98}},
  \bibinfo{pages}{094507} (\bibinfo{year}{2018}{\natexlab{a}}),
  \eprint{1807.06671}.

\bibitem[{\citenamefont{Detmold and Lin}(2006)}]{Detmold:2005gg}
\bibinfo{author}{\bibfnamefont{W.}~\bibnamefont{Detmold}} \bibnamefont{and}
  \bibinfo{author}{\bibfnamefont{C.~J.~D.} \bibnamefont{Lin}},
  \bibinfo{journal}{Phys. Rev. D} \textbf{\bibinfo{volume}{73}},
  \bibinfo{pages}{014501} (\bibinfo{year}{2006}), \eprint{hep-lat/0507007}.

\bibitem[{\citenamefont{Detmold et~al.}(2021)\citenamefont{Detmold, Grebe,
  Kanamori, Lin, Mondal, Perry, and Zhao}}]{Detmold:2021qln}
\bibinfo{author}{\bibfnamefont{W.}~\bibnamefont{Detmold}},
  \bibinfo{author}{\bibfnamefont{A.}~\bibnamefont{Grebe}},
  \bibinfo{author}{\bibfnamefont{I.}~\bibnamefont{Kanamori}},
  \bibinfo{author}{\bibfnamefont{C.~J.~D.} \bibnamefont{Lin}},
  \bibinfo{author}{\bibfnamefont{S.}~\bibnamefont{Mondal}},
  \bibinfo{author}{\bibfnamefont{R.}~\bibnamefont{Perry}}, \bibnamefont{and}
  \bibinfo{author}{\bibfnamefont{Y.}~\bibnamefont{Zhao}}
  (\bibinfo{year}{2021}), \eprint{2109.15241}.

\bibitem[{\citenamefont{Gao et~al.}(2022{\natexlab{a}})\citenamefont{Gao,
  Hanlon, Karthik, Mukherjee, Petreczky, Scior, Syritsyn, and
  Zhao}}]{Gao:2022vyh}
\bibinfo{author}{\bibfnamefont{X.}~\bibnamefont{Gao}},
  \bibinfo{author}{\bibfnamefont{A.~D.} \bibnamefont{Hanlon}},
  \bibinfo{author}{\bibfnamefont{N.}~\bibnamefont{Karthik}},
  \bibinfo{author}{\bibfnamefont{S.}~\bibnamefont{Mukherjee}},
  \bibinfo{author}{\bibfnamefont{P.}~\bibnamefont{Petreczky}},
  \bibinfo{author}{\bibfnamefont{P.}~\bibnamefont{Scior}},
  \bibinfo{author}{\bibfnamefont{S.}~\bibnamefont{Syritsyn}}, \bibnamefont{and}
  \bibinfo{author}{\bibfnamefont{Y.}~\bibnamefont{Zhao}},
  \bibinfo{journal}{Phys. Rev. D} \textbf{\bibinfo{volume}{106}},
  \bibinfo{pages}{074505} (\bibinfo{year}{2022}{\natexlab{a}}),
  \eprint{2206.04084}.

\bibitem[{\citenamefont{Blossier et~al.}(2024)\citenamefont{Blossier,
  Mangin-Brinet, Morgado~Ch\'avez, and San~Jos\'e}}]{Blossier:2024wyx}
\bibinfo{author}{\bibfnamefont{B.}~\bibnamefont{Blossier}},
  \bibinfo{author}{\bibfnamefont{M.}~\bibnamefont{Mangin-Brinet}},
  \bibinfo{author}{\bibfnamefont{J.~M.} \bibnamefont{Morgado~Ch\'avez}},
  \bibnamefont{and}
  \bibinfo{author}{\bibfnamefont{T.}~\bibnamefont{San~Jos\'e}}
  (\bibinfo{year}{2024}), \eprint{2406.04668}.

\bibitem[{\citenamefont{Ji}(2013)}]{Ji:2013dva}
\bibinfo{author}{\bibfnamefont{X.}~\bibnamefont{Ji}}, \bibinfo{journal}{Phys.
  Rev. Lett.} \textbf{\bibinfo{volume}{110}}, \bibinfo{pages}{262002}
  (\bibinfo{year}{2013}), \eprint{1305.1539}.

\bibitem[{\citenamefont{Ji}(2014)}]{Ji:2014gla}
\bibinfo{author}{\bibfnamefont{X.}~\bibnamefont{Ji}}, \bibinfo{journal}{Sci.
  China Phys. Mech. Astron.} \textbf{\bibinfo{volume}{57}},
  \bibinfo{pages}{1407} (\bibinfo{year}{2014}), \eprint{1404.6680}.

\bibitem[{\citenamefont{Ji et~al.}(2021{\natexlab{a}})\citenamefont{Ji, Liu,
  Liu, Zhang, and Zhao}}]{Ji:2020ect}
\bibinfo{author}{\bibfnamefont{X.}~\bibnamefont{Ji}},
  \bibinfo{author}{\bibfnamefont{Y.-S.} \bibnamefont{Liu}},
  \bibinfo{author}{\bibfnamefont{Y.}~\bibnamefont{Liu}},
  \bibinfo{author}{\bibfnamefont{J.-H.} \bibnamefont{Zhang}}, \bibnamefont{and}
  \bibinfo{author}{\bibfnamefont{Y.}~\bibnamefont{Zhao}},
  \bibinfo{journal}{Rev. Mod. Phys.} \textbf{\bibinfo{volume}{93}},
  \bibinfo{pages}{035005} (\bibinfo{year}{2021}{\natexlab{a}}),
  \eprint{2004.03543}.

\bibitem[{\citenamefont{Zhang et~al.}(2017)\citenamefont{Zhang, Chen, Ji, Jin,
  and Lin}}]{Zhang:2017bzy}
\bibinfo{author}{\bibfnamefont{J.-H.} \bibnamefont{Zhang}},
  \bibinfo{author}{\bibfnamefont{J.-W.} \bibnamefont{Chen}},
  \bibinfo{author}{\bibfnamefont{X.}~\bibnamefont{Ji}},
  \bibinfo{author}{\bibfnamefont{L.}~\bibnamefont{Jin}}, \bibnamefont{and}
  \bibinfo{author}{\bibfnamefont{H.-W.} \bibnamefont{Lin}},
  \bibinfo{journal}{Phys. Rev. D} \textbf{\bibinfo{volume}{95}},
  \bibinfo{pages}{094514} (\bibinfo{year}{2017}), \eprint{1702.00008}.

\bibitem[{\citenamefont{Zhang et~al.}(2019)\citenamefont{Zhang, Jin, Lin,
  Sch\"afer, Sun, Yang, Zhang, Zhao, and Chen}}]{Zhang:2017zfe}
\bibinfo{author}{\bibfnamefont{J.-H.} \bibnamefont{Zhang}},
  \bibinfo{author}{\bibfnamefont{L.}~\bibnamefont{Jin}},
  \bibinfo{author}{\bibfnamefont{H.-W.} \bibnamefont{Lin}},
  \bibinfo{author}{\bibfnamefont{A.}~\bibnamefont{Sch\"afer}},
  \bibinfo{author}{\bibfnamefont{P.}~\bibnamefont{Sun}},
  \bibinfo{author}{\bibfnamefont{Y.-B.} \bibnamefont{Yang}},
  \bibinfo{author}{\bibfnamefont{R.}~\bibnamefont{Zhang}},
  \bibinfo{author}{\bibfnamefont{Y.}~\bibnamefont{Zhao}}, \bibnamefont{and}
  \bibinfo{author}{\bibfnamefont{J.-W.} \bibnamefont{Chen}}
  (\bibinfo{collaboration}{LP3}), \bibinfo{journal}{Nucl. Phys. B}
  \textbf{\bibinfo{volume}{939}}, \bibinfo{pages}{429} (\bibinfo{year}{2019}),
  \eprint{1712.10025}.

\bibitem[{\citenamefont{Zhang et~al.}(2020)\citenamefont{Zhang, Honkala, Lin,
  and Chen}}]{Zhang:2020gaj}
\bibinfo{author}{\bibfnamefont{R.}~\bibnamefont{Zhang}},
  \bibinfo{author}{\bibfnamefont{C.}~\bibnamefont{Honkala}},
  \bibinfo{author}{\bibfnamefont{H.-W.} \bibnamefont{Lin}}, \bibnamefont{and}
  \bibinfo{author}{\bibfnamefont{J.-W.} \bibnamefont{Chen}},
  \bibinfo{journal}{Phys. Rev. D} \textbf{\bibinfo{volume}{102}},
  \bibinfo{pages}{094519} (\bibinfo{year}{2020}), \eprint{2005.13955}.

\bibitem[{\citenamefont{Hua et~al.}(2021)\citenamefont{Hua, Chu, Sun, Wang, Xu,
  Yang, Zhang, and Zhang}}]{Hua:2020gnw}
\bibinfo{author}{\bibfnamefont{J.}~\bibnamefont{Hua}},
  \bibinfo{author}{\bibfnamefont{M.-H.} \bibnamefont{Chu}},
  \bibinfo{author}{\bibfnamefont{P.}~\bibnamefont{Sun}},
  \bibinfo{author}{\bibfnamefont{W.}~\bibnamefont{Wang}},
  \bibinfo{author}{\bibfnamefont{J.}~\bibnamefont{Xu}},
  \bibinfo{author}{\bibfnamefont{Y.-B.} \bibnamefont{Yang}},
  \bibinfo{author}{\bibfnamefont{J.-H.} \bibnamefont{Zhang}}, \bibnamefont{and}
  \bibinfo{author}{\bibfnamefont{Q.-A.} \bibnamefont{Zhang}}
  (\bibinfo{collaboration}{Lattice Parton}), \bibinfo{journal}{Phys. Rev.
  Lett.} \textbf{\bibinfo{volume}{127}}, \bibinfo{pages}{062002}
  (\bibinfo{year}{2021}), \eprint{2011.09788}.

\bibitem[{\citenamefont{Hua et~al.}(2022)}]{LatticeParton:2022zqc}
\bibinfo{author}{\bibfnamefont{J.}~\bibnamefont{Hua}} \bibnamefont{et~al.}
  (\bibinfo{collaboration}{Lattice Parton}), \bibinfo{journal}{Phys. Rev.
  Lett.} \textbf{\bibinfo{volume}{129}}, \bibinfo{pages}{132001}
  (\bibinfo{year}{2022}), \eprint{2201.09173}.

\bibitem[{\citenamefont{Holligan et~al.}(2023)\citenamefont{Holligan, Ji, Lin,
  Su, and Zhang}}]{Holligan:2023rex}
\bibinfo{author}{\bibfnamefont{J.}~\bibnamefont{Holligan}},
  \bibinfo{author}{\bibfnamefont{X.}~\bibnamefont{Ji}},
  \bibinfo{author}{\bibfnamefont{H.-W.} \bibnamefont{Lin}},
  \bibinfo{author}{\bibfnamefont{Y.}~\bibnamefont{Su}}, \bibnamefont{and}
  \bibinfo{author}{\bibfnamefont{R.}~\bibnamefont{Zhang}},
  \bibinfo{journal}{Nucl. Phys. B} \textbf{\bibinfo{volume}{993}},
  \bibinfo{pages}{116282} (\bibinfo{year}{2023}), \eprint{2301.10372}.

\bibitem[{\citenamefont{Baker et~al.}(2024)\citenamefont{Baker, Bollweg, Boyle,
  Clo\"et, Gao, Mukherjee, Petreczky, Zhang, and Zhao}}]{Baker:2024zcd}
\bibinfo{author}{\bibfnamefont{E.}~\bibnamefont{Baker}},
  \bibinfo{author}{\bibfnamefont{D.}~\bibnamefont{Bollweg}},
  \bibinfo{author}{\bibfnamefont{P.}~\bibnamefont{Boyle}},
  \bibinfo{author}{\bibfnamefont{I.}~\bibnamefont{Clo\"et}},
  \bibinfo{author}{\bibfnamefont{X.}~\bibnamefont{Gao}},
  \bibinfo{author}{\bibfnamefont{S.}~\bibnamefont{Mukherjee}},
  \bibinfo{author}{\bibfnamefont{P.}~\bibnamefont{Petreczky}},
  \bibinfo{author}{\bibfnamefont{R.}~\bibnamefont{Zhang}}, \bibnamefont{and}
  \bibinfo{author}{\bibfnamefont{Y.}~\bibnamefont{Zhao}}
  (\bibinfo{year}{2024}), \eprint{2405.20120}.

\bibitem[{\citenamefont{Chen et~al.}(2017)\citenamefont{Chen, Ji, and
  Zhang}}]{Chen:2016fxx}
\bibinfo{author}{\bibfnamefont{J.-W.} \bibnamefont{Chen}},
  \bibinfo{author}{\bibfnamefont{X.}~\bibnamefont{Ji}}, \bibnamefont{and}
  \bibinfo{author}{\bibfnamefont{J.-H.} \bibnamefont{Zhang}},
  \bibinfo{journal}{Nucl. Phys. B} \textbf{\bibinfo{volume}{915}},
  \bibinfo{pages}{1} (\bibinfo{year}{2017}), \eprint{1609.08102}.

\bibitem[{\citenamefont{Ji et~al.}(2018)\citenamefont{Ji, Zhang, and
  Zhao}}]{Ji:2017oey}
\bibinfo{author}{\bibfnamefont{X.}~\bibnamefont{Ji}},
  \bibinfo{author}{\bibfnamefont{J.-H.} \bibnamefont{Zhang}}, \bibnamefont{and}
  \bibinfo{author}{\bibfnamefont{Y.}~\bibnamefont{Zhao}},
  \bibinfo{journal}{Phys. Rev. Lett.} \textbf{\bibinfo{volume}{120}},
  \bibinfo{pages}{112001} (\bibinfo{year}{2018}), \eprint{1706.08962}.

\bibitem[{\citenamefont{Green et~al.}(2018)\citenamefont{Green, Jansen, and
  Steffens}}]{Green:2017xeu}
\bibinfo{author}{\bibfnamefont{J.}~\bibnamefont{Green}},
  \bibinfo{author}{\bibfnamefont{K.}~\bibnamefont{Jansen}}, \bibnamefont{and}
  \bibinfo{author}{\bibfnamefont{F.}~\bibnamefont{Steffens}},
  \bibinfo{journal}{Phys. Rev. Lett.} \textbf{\bibinfo{volume}{121}},
  \bibinfo{pages}{022004} (\bibinfo{year}{2018}), \eprint{1707.07152}.

\bibitem[{\citenamefont{Ishikawa et~al.}(2017)\citenamefont{Ishikawa, Ma, Qiu,
  and Yoshida}}]{Ishikawa:2017faj}
\bibinfo{author}{\bibfnamefont{T.}~\bibnamefont{Ishikawa}},
  \bibinfo{author}{\bibfnamefont{Y.-Q.} \bibnamefont{Ma}},
  \bibinfo{author}{\bibfnamefont{J.-W.} \bibnamefont{Qiu}}, \bibnamefont{and}
  \bibinfo{author}{\bibfnamefont{S.}~\bibnamefont{Yoshida}},
  \bibinfo{journal}{Phys. Rev.} \textbf{\bibinfo{volume}{D96}},
  \bibinfo{pages}{094019} (\bibinfo{year}{2017}), \eprint{1707.03107}.

\bibitem[{\citenamefont{Constantinou and
  Panagopoulos}(2017)}]{Constantinou:2017sej}
\bibinfo{author}{\bibfnamefont{M.}~\bibnamefont{Constantinou}}
  \bibnamefont{and}
  \bibinfo{author}{\bibfnamefont{H.}~\bibnamefont{Panagopoulos}},
  \bibinfo{journal}{Phys. Rev. D} \textbf{\bibinfo{volume}{96}},
  \bibinfo{pages}{054506} (\bibinfo{year}{2017}), \eprint{1705.11193}.

\bibitem[{\citenamefont{Alexandrou et~al.}(2017)\citenamefont{Alexandrou,
  Cichy, Constantinou, Hadjiyiannakou, Jansen, Panagopoulos, and
  Steffens}}]{Alexandrou_2017}
\bibinfo{author}{\bibfnamefont{C.}~\bibnamefont{Alexandrou}},
  \bibinfo{author}{\bibfnamefont{K.}~\bibnamefont{Cichy}},
  \bibinfo{author}{\bibfnamefont{M.}~\bibnamefont{Constantinou}},
  \bibinfo{author}{\bibfnamefont{K.}~\bibnamefont{Hadjiyiannakou}},
  \bibinfo{author}{\bibfnamefont{K.}~\bibnamefont{Jansen}},
  \bibinfo{author}{\bibfnamefont{H.}~\bibnamefont{Panagopoulos}},
  \bibnamefont{and} \bibinfo{author}{\bibfnamefont{F.}~\bibnamefont{Steffens}},
  \bibinfo{journal}{Nuclear Physics B} \textbf{\bibinfo{volume}{923}},
  \bibinfo{pages}{394} (\bibinfo{year}{2017}).

\bibitem[{\citenamefont{Chen et~al.}(2018)\citenamefont{Chen, Ishikawa, Jin,
  Lin, Yang, Zhang, and Zhao}}]{Chen:2017mzz}
\bibinfo{author}{\bibfnamefont{J.-W.} \bibnamefont{Chen}},
  \bibinfo{author}{\bibfnamefont{T.}~\bibnamefont{Ishikawa}},
  \bibinfo{author}{\bibfnamefont{L.}~\bibnamefont{Jin}},
  \bibinfo{author}{\bibfnamefont{H.-W.} \bibnamefont{Lin}},
  \bibinfo{author}{\bibfnamefont{Y.-B.} \bibnamefont{Yang}},
  \bibinfo{author}{\bibfnamefont{J.-H.} \bibnamefont{Zhang}}, \bibnamefont{and}
  \bibinfo{author}{\bibfnamefont{Y.}~\bibnamefont{Zhao}},
  \bibinfo{journal}{Phys. Rev.} \textbf{\bibinfo{volume}{D97}},
  \bibinfo{pages}{014505} (\bibinfo{year}{2018}), \eprint{1706.01295}.

\bibitem[{\citenamefont{Stewart and Zhao}(2018)}]{Stewart:2017tvs}
\bibinfo{author}{\bibfnamefont{I.~W.} \bibnamefont{Stewart}} \bibnamefont{and}
  \bibinfo{author}{\bibfnamefont{Y.}~\bibnamefont{Zhao}},
  \bibinfo{journal}{Phys. Rev. D} \textbf{\bibinfo{volume}{97}},
  \bibinfo{pages}{054512} (\bibinfo{year}{2018}), \eprint{1709.04933}.

\bibitem[{\citenamefont{Orginos et~al.}(2017)\citenamefont{Orginos, Radyushkin,
  Karpie, and Zafeiropoulos}}]{Orginos:2017kos}
\bibinfo{author}{\bibfnamefont{K.}~\bibnamefont{Orginos}},
  \bibinfo{author}{\bibfnamefont{A.}~\bibnamefont{Radyushkin}},
  \bibinfo{author}{\bibfnamefont{J.}~\bibnamefont{Karpie}}, \bibnamefont{and}
  \bibinfo{author}{\bibfnamefont{S.}~\bibnamefont{Zafeiropoulos}},
  \bibinfo{journal}{Phys. Rev. D} \textbf{\bibinfo{volume}{96}},
  \bibinfo{pages}{094503} (\bibinfo{year}{2017}), \eprint{1706.05373}.

\bibitem[{\citenamefont{Fan et~al.}(2020)\citenamefont{Fan, Gao, Li, Lin,
  Karthik, Mukherjee, Petreczky, Syritsyn, Yang, and Zhang}}]{Fan:2020nzz}
\bibinfo{author}{\bibfnamefont{Z.}~\bibnamefont{Fan}},
  \bibinfo{author}{\bibfnamefont{X.}~\bibnamefont{Gao}},
  \bibinfo{author}{\bibfnamefont{R.}~\bibnamefont{Li}},
  \bibinfo{author}{\bibfnamefont{H.-W.} \bibnamefont{Lin}},
  \bibinfo{author}{\bibfnamefont{N.}~\bibnamefont{Karthik}},
  \bibinfo{author}{\bibfnamefont{S.}~\bibnamefont{Mukherjee}},
  \bibinfo{author}{\bibfnamefont{P.}~\bibnamefont{Petreczky}},
  \bibinfo{author}{\bibfnamefont{S.}~\bibnamefont{Syritsyn}},
  \bibinfo{author}{\bibfnamefont{Y.-B.} \bibnamefont{Yang}}, \bibnamefont{and}
  \bibinfo{author}{\bibfnamefont{R.}~\bibnamefont{Zhang}},
  \bibinfo{journal}{Phys. Rev. D} \textbf{\bibinfo{volume}{102}},
  \bibinfo{pages}{074504} (\bibinfo{year}{2020}), \eprint{2005.12015}.

\bibitem[{\citenamefont{Ji et~al.}(2021{\natexlab{b}})\citenamefont{Ji, Liu,
  Sch\"afer, Wang, Yang, Zhang, and Zhao}}]{Ji:2020brr}
\bibinfo{author}{\bibfnamefont{X.}~\bibnamefont{Ji}},
  \bibinfo{author}{\bibfnamefont{Y.}~\bibnamefont{Liu}},
  \bibinfo{author}{\bibfnamefont{A.}~\bibnamefont{Sch\"afer}},
  \bibinfo{author}{\bibfnamefont{W.}~\bibnamefont{Wang}},
  \bibinfo{author}{\bibfnamefont{Y.-B.} \bibnamefont{Yang}},
  \bibinfo{author}{\bibfnamefont{J.-H.} \bibnamefont{Zhang}}, \bibnamefont{and}
  \bibinfo{author}{\bibfnamefont{Y.}~\bibnamefont{Zhao}},
  \bibinfo{journal}{Nucl. Phys. B} \textbf{\bibinfo{volume}{964}},
  \bibinfo{pages}{115311} (\bibinfo{year}{2021}{\natexlab{b}}),
  \eprint{2008.03886}.

\bibitem[{\citenamefont{Huo
  et~al.}(2021)}]{LatticePartonCollaborationLPC:2021xdx}
\bibinfo{author}{\bibfnamefont{Y.-K.} \bibnamefont{Huo}} \bibnamefont{et~al.}
  (\bibinfo{collaboration}{Lattice Parton Collaboration (LPC)}),
  \bibinfo{journal}{Nucl. Phys. B} \textbf{\bibinfo{volume}{969}},
  \bibinfo{pages}{115443} (\bibinfo{year}{2021}), \eprint{2103.02965}.

\bibitem[{\citenamefont{Gao et~al.}(2022{\natexlab{b}})\citenamefont{Gao,
  Hanlon, Mukherjee, Petreczky, Scior, Syritsyn, and Zhao}}]{Gao:2021dbh}
\bibinfo{author}{\bibfnamefont{X.}~\bibnamefont{Gao}},
  \bibinfo{author}{\bibfnamefont{A.~D.} \bibnamefont{Hanlon}},
  \bibinfo{author}{\bibfnamefont{S.}~\bibnamefont{Mukherjee}},
  \bibinfo{author}{\bibfnamefont{P.}~\bibnamefont{Petreczky}},
  \bibinfo{author}{\bibfnamefont{P.}~\bibnamefont{Scior}},
  \bibinfo{author}{\bibfnamefont{S.}~\bibnamefont{Syritsyn}}, \bibnamefont{and}
  \bibinfo{author}{\bibfnamefont{Y.}~\bibnamefont{Zhao}},
  \bibinfo{journal}{Phys. Rev. Lett.} \textbf{\bibinfo{volume}{128}},
  \bibinfo{pages}{142003} (\bibinfo{year}{2022}{\natexlab{b}}),
  \eprint{2112.02208}.

\bibitem[{\citenamefont{Zhang et~al.}(2023)\citenamefont{Zhang, Holligan, Ji,
  and Su}}]{Zhang:2023bxs}
\bibinfo{author}{\bibfnamefont{R.}~\bibnamefont{Zhang}},
  \bibinfo{author}{\bibfnamefont{J.}~\bibnamefont{Holligan}},
  \bibinfo{author}{\bibfnamefont{X.}~\bibnamefont{Ji}}, \bibnamefont{and}
  \bibinfo{author}{\bibfnamefont{Y.}~\bibnamefont{Su}}, \bibinfo{journal}{Phys.
  Lett. B} \textbf{\bibinfo{volume}{844}}, \bibinfo{pages}{138081}
  (\bibinfo{year}{2023}), \eprint{2305.05212}.

\bibitem[{\citenamefont{Bali et~al.}(2018{\natexlab{b}})}]{Bali:2017gfr}
\bibinfo{author}{\bibfnamefont{G.~S.} \bibnamefont{Bali}} \bibnamefont{et~al.},
  \bibinfo{journal}{Eur. Phys. J. C} \textbf{\bibinfo{volume}{78}},
  \bibinfo{pages}{217} (\bibinfo{year}{2018}{\natexlab{b}}),
  \eprint{1709.04325}.

\bibitem[{\citenamefont{Xu et~al.}(2018)\citenamefont{Xu, Zhang, and
  Zhao}}]{Xu:2018mpf}
\bibinfo{author}{\bibfnamefont{J.}~\bibnamefont{Xu}},
  \bibinfo{author}{\bibfnamefont{Q.-A.} \bibnamefont{Zhang}}, \bibnamefont{and}
  \bibinfo{author}{\bibfnamefont{S.}~\bibnamefont{Zhao}},
  \bibinfo{journal}{Phys. Rev. D} \textbf{\bibinfo{volume}{97}},
  \bibinfo{pages}{114026} (\bibinfo{year}{2018}), \eprint{1804.01042}.

\bibitem[{\citenamefont{Liu et~al.}(2019)\citenamefont{Liu, Wang, Xu, Zhang,
  Zhao, and Zhao}}]{Liu:2018tox}
\bibinfo{author}{\bibfnamefont{Y.-S.} \bibnamefont{Liu}},
  \bibinfo{author}{\bibfnamefont{W.}~\bibnamefont{Wang}},
  \bibinfo{author}{\bibfnamefont{J.}~\bibnamefont{Xu}},
  \bibinfo{author}{\bibfnamefont{Q.-A.} \bibnamefont{Zhang}},
  \bibinfo{author}{\bibfnamefont{S.}~\bibnamefont{Zhao}}, \bibnamefont{and}
  \bibinfo{author}{\bibfnamefont{Y.}~\bibnamefont{Zhao}},
  \bibinfo{journal}{Phys. Rev. D} \textbf{\bibinfo{volume}{99}},
  \bibinfo{pages}{094036} (\bibinfo{year}{2019}), \eprint{1810.10879}.

\bibitem[{\citenamefont{Gao et~al.}(2021)\citenamefont{Gao, Lee, Mukherjee,
  Shugert, and Zhao}}]{Gao:2021hxl}
\bibinfo{author}{\bibfnamefont{X.}~\bibnamefont{Gao}},
  \bibinfo{author}{\bibfnamefont{K.}~\bibnamefont{Lee}},
  \bibinfo{author}{\bibfnamefont{S.}~\bibnamefont{Mukherjee}},
  \bibinfo{author}{\bibfnamefont{C.}~\bibnamefont{Shugert}}, \bibnamefont{and}
  \bibinfo{author}{\bibfnamefont{Y.}~\bibnamefont{Zhao}},
  \bibinfo{journal}{Phys. Rev. D} \textbf{\bibinfo{volume}{103}},
  \bibinfo{pages}{094504} (\bibinfo{year}{2021}), \eprint{2102.01101}.

\bibitem[{\citenamefont{Su et~al.}(2022)\citenamefont{Su, Holligan, Ji, Yao,
  Zhang, and Zhang}}]{Su:2022fiu}
\bibinfo{author}{\bibfnamefont{Y.}~\bibnamefont{Su}},
  \bibinfo{author}{\bibfnamefont{J.}~\bibnamefont{Holligan}},
  \bibinfo{author}{\bibfnamefont{X.}~\bibnamefont{Ji}},
  \bibinfo{author}{\bibfnamefont{F.}~\bibnamefont{Yao}},
  \bibinfo{author}{\bibfnamefont{J.-H.} \bibnamefont{Zhang}}, \bibnamefont{and}
  \bibinfo{author}{\bibfnamefont{R.}~\bibnamefont{Zhang}}
  (\bibinfo{year}{2022}), \eprint{2209.01236}.

\bibitem[{\citenamefont{Ji et~al.}(2023)\citenamefont{Ji, Liu, and
  Su}}]{Ji:2023pba}
\bibinfo{author}{\bibfnamefont{X.}~\bibnamefont{Ji}},
  \bibinfo{author}{\bibfnamefont{Y.}~\bibnamefont{Liu}}, \bibnamefont{and}
  \bibinfo{author}{\bibfnamefont{Y.}~\bibnamefont{Su}}, \bibinfo{journal}{JHEP}
  \textbf{\bibinfo{volume}{08}}, \bibinfo{pages}{037} (\bibinfo{year}{2023}),
  \eprint{2305.04416}.

\bibitem[{\citenamefont{Liu and Su}(2023)}]{Liu:2023onm}
\bibinfo{author}{\bibfnamefont{Y.}~\bibnamefont{Liu}} \bibnamefont{and}
  \bibinfo{author}{\bibfnamefont{Y.}~\bibnamefont{Su}} (\bibinfo{year}{2023}),
  \eprint{2311.06907}.

\bibitem[{\citenamefont{Ji}(2022)}]{Ji:2022ezo}
\bibinfo{author}{\bibfnamefont{X.}~\bibnamefont{Ji}} (\bibinfo{year}{2022}),
  \eprint{2209.09332}.

\bibitem[{\citenamefont{Follana et~al.}(2007)\citenamefont{Follana, Mason,
  Davies, Hornbostel, Lepage, Shigemitsu, Trottier, and Wong}}]{Follana:2006rc}
\bibinfo{author}{\bibfnamefont{E.}~\bibnamefont{Follana}},
  \bibinfo{author}{\bibfnamefont{Q.}~\bibnamefont{Mason}},
  \bibinfo{author}{\bibfnamefont{C.}~\bibnamefont{Davies}},
  \bibinfo{author}{\bibfnamefont{K.}~\bibnamefont{Hornbostel}},
  \bibinfo{author}{\bibfnamefont{G.~P.} \bibnamefont{Lepage}},
  \bibinfo{author}{\bibfnamefont{J.}~\bibnamefont{Shigemitsu}},
  \bibinfo{author}{\bibfnamefont{H.}~\bibnamefont{Trottier}}, \bibnamefont{and}
  \bibinfo{author}{\bibfnamefont{K.}~\bibnamefont{Wong}}
  (\bibinfo{collaboration}{HPQCD, UKQCD}), \bibinfo{journal}{Phys. Rev. D}
  \textbf{\bibinfo{volume}{75}}, \bibinfo{pages}{054502}
  (\bibinfo{year}{2007}), \eprint{hep-lat/0610092}.

\bibitem[{\citenamefont{Bazavov et~al.}(2019)}]{Bazavov:2019www}
\bibinfo{author}{\bibfnamefont{A.}~\bibnamefont{Bazavov}} \bibnamefont{et~al.},
  \bibinfo{journal}{Phys. Rev. D} \textbf{\bibinfo{volume}{100}},
  \bibinfo{pages}{094510} (\bibinfo{year}{2019}), \eprint{1908.09552}.

\bibitem[{\citenamefont{Gao et~al.}(2022{\natexlab{c}})\citenamefont{Gao,
  Hanlon, Karthik, Mukherjee, Petreczky, Scior, Shi, Syritsyn, Zhao, and
  Zhou}}]{Gao:2022iex}
\bibinfo{author}{\bibfnamefont{X.}~\bibnamefont{Gao}},
  \bibinfo{author}{\bibfnamefont{A.~D.} \bibnamefont{Hanlon}},
  \bibinfo{author}{\bibfnamefont{N.}~\bibnamefont{Karthik}},
  \bibinfo{author}{\bibfnamefont{S.}~\bibnamefont{Mukherjee}},
  \bibinfo{author}{\bibfnamefont{P.}~\bibnamefont{Petreczky}},
  \bibinfo{author}{\bibfnamefont{P.}~\bibnamefont{Scior}},
  \bibinfo{author}{\bibfnamefont{S.}~\bibnamefont{Shi}},
  \bibinfo{author}{\bibfnamefont{S.}~\bibnamefont{Syritsyn}},
  \bibinfo{author}{\bibfnamefont{Y.}~\bibnamefont{Zhao}}, \bibnamefont{and}
  \bibinfo{author}{\bibfnamefont{K.}~\bibnamefont{Zhou}},
  \bibinfo{journal}{Phys. Rev. D} \textbf{\bibinfo{volume}{106}},
  \bibinfo{pages}{114510} (\bibinfo{year}{2022}{\natexlab{c}}),
  \eprint{2208.02297}.

\bibitem[{\citenamefont{Hasenfratz and Knechtli}(2001)}]{Hasenfratz:2001hp}
\bibinfo{author}{\bibfnamefont{A.}~\bibnamefont{Hasenfratz}} \bibnamefont{and}
  \bibinfo{author}{\bibfnamefont{F.}~\bibnamefont{Knechtli}},
  \bibinfo{journal}{Phys. Rev. D} \textbf{\bibinfo{volume}{64}},
  \bibinfo{pages}{034504} (\bibinfo{year}{2001}), \eprint{hep-lat/0103029}.

\bibitem[{\citenamefont{Brannick et~al.}(2008)\citenamefont{Brannick, Brower,
  Clark, Osborn, and Rebbi}}]{Brannick:2007ue}
\bibinfo{author}{\bibfnamefont{J.}~\bibnamefont{Brannick}},
  \bibinfo{author}{\bibfnamefont{R.~C.} \bibnamefont{Brower}},
  \bibinfo{author}{\bibfnamefont{M.~A.} \bibnamefont{Clark}},
  \bibinfo{author}{\bibfnamefont{J.~C.} \bibnamefont{Osborn}},
  \bibnamefont{and} \bibinfo{author}{\bibfnamefont{C.}~\bibnamefont{Rebbi}},
  \bibinfo{journal}{Phys. Rev. Lett.} \textbf{\bibinfo{volume}{100}},
  \bibinfo{pages}{041601} (\bibinfo{year}{2008}), \eprint{0707.4018}.

\bibitem[{\citenamefont{Clark et~al.}(2010)\citenamefont{Clark, Babich, Barros,
  Brower, and Rebbi}}]{Clark:2009wm}
\bibinfo{author}{\bibfnamefont{M.~A.} \bibnamefont{Clark}},
  \bibinfo{author}{\bibfnamefont{R.}~\bibnamefont{Babich}},
  \bibinfo{author}{\bibfnamefont{K.}~\bibnamefont{Barros}},
  \bibinfo{author}{\bibfnamefont{R.~C.} \bibnamefont{Brower}},
  \bibnamefont{and} \bibinfo{author}{\bibfnamefont{C.}~\bibnamefont{Rebbi}},
  \bibinfo{journal}{Comput. Phys. Commun.} \textbf{\bibinfo{volume}{181}},
  \bibinfo{pages}{1517} (\bibinfo{year}{2010}), \eprint{0911.3191}.

\bibitem[{\citenamefont{Babich et~al.}(2011)\citenamefont{Babich, Clark, Joo,
  Shi, Brower, and Gottlieb}}]{Babich:2011np}
\bibinfo{author}{\bibfnamefont{R.}~\bibnamefont{Babich}},
  \bibinfo{author}{\bibfnamefont{M.~A.} \bibnamefont{Clark}},
  \bibinfo{author}{\bibfnamefont{B.}~\bibnamefont{Joo}},
  \bibinfo{author}{\bibfnamefont{G.}~\bibnamefont{Shi}},
  \bibinfo{author}{\bibfnamefont{R.~C.} \bibnamefont{Brower}},
  \bibnamefont{and} \bibinfo{author}{\bibfnamefont{S.}~\bibnamefont{Gottlieb}},
  in \emph{\bibinfo{booktitle}{{SC11 International Conference for High
  Performance Computing, Networking, Storage and Analysis Seattle, Washington,
  November 12-18, 2011}}} (\bibinfo{year}{2011}), \eprint{1109.2935}.

\bibitem[{\citenamefont{Clark et~al.}(2016)\citenamefont{Clark, Jo,
  Strelchenko, Cheng, Gambhir, and Brower}}]{Clark:2016rdz}
\bibinfo{author}{\bibfnamefont{M.~A.} \bibnamefont{Clark}},
  \bibinfo{author}{\bibfnamefont{B.}~\bibnamefont{Jo}},
  \bibinfo{author}{\bibfnamefont{A.}~\bibnamefont{Strelchenko}},
  \bibinfo{author}{\bibfnamefont{M.}~\bibnamefont{Cheng}},
  \bibinfo{author}{\bibfnamefont{A.}~\bibnamefont{Gambhir}}, \bibnamefont{and}
  \bibinfo{author}{\bibfnamefont{R.}~\bibnamefont{Brower}}
  (\bibinfo{year}{2016}), \eprint{1612.07873}.

\bibitem[{\citenamefont{Shintani et~al.}(2015)\citenamefont{Shintani, Arthur,
  Blum, Izubuchi, Jung, and Lehner}}]{Shintani:2014vja}
\bibinfo{author}{\bibfnamefont{E.}~\bibnamefont{Shintani}},
  \bibinfo{author}{\bibfnamefont{R.}~\bibnamefont{Arthur}},
  \bibinfo{author}{\bibfnamefont{T.}~\bibnamefont{Blum}},
  \bibinfo{author}{\bibfnamefont{T.}~\bibnamefont{Izubuchi}},
  \bibinfo{author}{\bibfnamefont{C.}~\bibnamefont{Jung}}, \bibnamefont{and}
  \bibinfo{author}{\bibfnamefont{C.}~\bibnamefont{Lehner}},
  \bibinfo{journal}{Phys. Rev. D} \textbf{\bibinfo{volume}{91}},
  \bibinfo{pages}{114511} (\bibinfo{year}{2015}), \eprint{1402.0244}.

\bibitem[{\citenamefont{Bali et~al.}(2016)\citenamefont{Bali, Lang, Musch, and
  Sch\"afer}}]{Bali:2016lva}
\bibinfo{author}{\bibfnamefont{G.~S.} \bibnamefont{Bali}},
  \bibinfo{author}{\bibfnamefont{B.}~\bibnamefont{Lang}},
  \bibinfo{author}{\bibfnamefont{B.~U.} \bibnamefont{Musch}}, \bibnamefont{and}
  \bibinfo{author}{\bibfnamefont{A.}~\bibnamefont{Sch\"afer}},
  \bibinfo{journal}{Phys. Rev. D} \textbf{\bibinfo{volume}{93}},
  \bibinfo{pages}{094515} (\bibinfo{year}{2016}), \eprint{1602.05525}.

\bibitem[{\citenamefont{Izubuchi et~al.}(2019)\citenamefont{Izubuchi, Jin,
  Kallidonis, Karthik, Mukherjee, Petreczky, Shugert, and
  Syritsyn}}]{Izubuchi:2019lyk}
\bibinfo{author}{\bibfnamefont{T.}~\bibnamefont{Izubuchi}},
  \bibinfo{author}{\bibfnamefont{L.}~\bibnamefont{Jin}},
  \bibinfo{author}{\bibfnamefont{C.}~\bibnamefont{Kallidonis}},
  \bibinfo{author}{\bibfnamefont{N.}~\bibnamefont{Karthik}},
  \bibinfo{author}{\bibfnamefont{S.}~\bibnamefont{Mukherjee}},
  \bibinfo{author}{\bibfnamefont{P.}~\bibnamefont{Petreczky}},
  \bibinfo{author}{\bibfnamefont{C.}~\bibnamefont{Shugert}}, \bibnamefont{and}
  \bibinfo{author}{\bibfnamefont{S.}~\bibnamefont{Syritsyn}},
  \bibinfo{journal}{Phys. Rev. D} \textbf{\bibinfo{volume}{100}},
  \bibinfo{pages}{034516} (\bibinfo{year}{2019}), \eprint{1905.06349}.

\bibitem[{\citenamefont{Chen et~al.}(2019)\citenamefont{Chen, Ishikawa, Jin,
  Lin, Zhang, and Zhao}}]{Chen:2017mie}
\bibinfo{author}{\bibfnamefont{J.-W.} \bibnamefont{Chen}},
  \bibinfo{author}{\bibfnamefont{T.}~\bibnamefont{Ishikawa}},
  \bibinfo{author}{\bibfnamefont{L.}~\bibnamefont{Jin}},
  \bibinfo{author}{\bibfnamefont{H.-W.} \bibnamefont{Lin}},
  \bibinfo{author}{\bibfnamefont{J.-H.} \bibnamefont{Zhang}}, \bibnamefont{and}
  \bibinfo{author}{\bibfnamefont{Y.}~\bibnamefont{Zhao}}
  (\bibinfo{collaboration}{LP3}), \bibinfo{journal}{Chin. Phys. C}
  \textbf{\bibinfo{volume}{43}}, \bibinfo{pages}{103101}
  (\bibinfo{year}{2019}), \eprint{1710.01089}.

\bibitem[{\citenamefont{Ding et~al.}(2024)\citenamefont{Ding, Gao, Hanlon,
  Mukherjee, Petreczky, Shi, Syritsyn, Zhang, and Zhao}}]{Ding:2024lfj}
\bibinfo{author}{\bibfnamefont{H.-T.} \bibnamefont{Ding}},
  \bibinfo{author}{\bibfnamefont{X.}~\bibnamefont{Gao}},
  \bibinfo{author}{\bibfnamefont{A.~D.} \bibnamefont{Hanlon}},
  \bibinfo{author}{\bibfnamefont{S.}~\bibnamefont{Mukherjee}},
  \bibinfo{author}{\bibfnamefont{P.}~\bibnamefont{Petreczky}},
  \bibinfo{author}{\bibfnamefont{Q.}~\bibnamefont{Shi}},
  \bibinfo{author}{\bibfnamefont{S.}~\bibnamefont{Syritsyn}},
  \bibinfo{author}{\bibfnamefont{R.}~\bibnamefont{Zhang}}, \bibnamefont{and}
  \bibinfo{author}{\bibfnamefont{Y.}~\bibnamefont{Zhao}}
  (\bibinfo{year}{2024}), \eprint{2404.04412}.

\bibitem[{\citenamefont{Bhattacharya et~al.}(2022)\citenamefont{Bhattacharya,
  Cichy, Constantinou, Dodson, Gao, Metz, Mukherjee, Scapellato, Steffens, and
  Zhao}}]{Bhattacharya:2022aob}
\bibinfo{author}{\bibfnamefont{S.}~\bibnamefont{Bhattacharya}},
  \bibinfo{author}{\bibfnamefont{K.}~\bibnamefont{Cichy}},
  \bibinfo{author}{\bibfnamefont{M.}~\bibnamefont{Constantinou}},
  \bibinfo{author}{\bibfnamefont{J.}~\bibnamefont{Dodson}},
  \bibinfo{author}{\bibfnamefont{X.}~\bibnamefont{Gao}},
  \bibinfo{author}{\bibfnamefont{A.}~\bibnamefont{Metz}},
  \bibinfo{author}{\bibfnamefont{S.}~\bibnamefont{Mukherjee}},
  \bibinfo{author}{\bibfnamefont{A.}~\bibnamefont{Scapellato}},
  \bibinfo{author}{\bibfnamefont{F.}~\bibnamefont{Steffens}}, \bibnamefont{and}
  \bibinfo{author}{\bibfnamefont{Y.}~\bibnamefont{Zhao}},
  \bibinfo{journal}{Phys. Rev. D} \textbf{\bibinfo{volume}{106}},
  \bibinfo{pages}{114512} (\bibinfo{year}{2022}), \eprint{2209.05373}.

\bibitem[{\citenamefont{Aoki
  et~al.}(2022)}]{FlavourLatticeAveragingGroupFLAG:2021npn}
\bibinfo{author}{\bibfnamefont{Y.}~\bibnamefont{Aoki}} \bibnamefont{et~al.}
  (\bibinfo{collaboration}{Flavour Lattice Averaging Group (FLAG)}),
  \bibinfo{journal}{Eur. Phys. J. C} \textbf{\bibinfo{volume}{82}},
  \bibinfo{pages}{869} (\bibinfo{year}{2022}), \eprint{2111.09849}.

\bibitem[{\citenamefont{Radyushkin}(2019)}]{Radyushkin:2019owq}
\bibinfo{author}{\bibfnamefont{A.~V.} \bibnamefont{Radyushkin}},
  \bibinfo{journal}{Phys. Rev. D} \textbf{\bibinfo{volume}{100}},
  \bibinfo{pages}{116011} (\bibinfo{year}{2019}), \eprint{1909.08474}.

\bibitem[{\citenamefont{Lepage and Brodsky}(1980)}]{Lepage:1980fj}
\bibinfo{author}{\bibfnamefont{G.~P.} \bibnamefont{Lepage}} \bibnamefont{and}
  \bibinfo{author}{\bibfnamefont{S.~J.} \bibnamefont{Brodsky}},
  \bibinfo{journal}{Phys. Rev. D} \textbf{\bibinfo{volume}{22}},
  \bibinfo{pages}{2157} (\bibinfo{year}{1980}).

\bibitem[{\citenamefont{Braun et~al.}(2017)\citenamefont{Braun, Manashov, Moch,
  and Strohmaier}}]{Braun:2017cih}
\bibinfo{author}{\bibfnamefont{V.~M.} \bibnamefont{Braun}},
  \bibinfo{author}{\bibfnamefont{A.~N.} \bibnamefont{Manashov}},
  \bibinfo{author}{\bibfnamefont{S.}~\bibnamefont{Moch}}, \bibnamefont{and}
  \bibinfo{author}{\bibfnamefont{M.}~\bibnamefont{Strohmaier}},
  \bibinfo{journal}{JHEP} \textbf{\bibinfo{volume}{06}}, \bibinfo{pages}{037}
  (\bibinfo{year}{2017}), \eprint{1703.09532}.

\bibitem[{\citenamefont{Donnellan et~al.}(2007)\citenamefont{Donnellan, Flynn,
  Juttner, Sachrajda, Antonio, Boyle, Maynard, Pendleton, and
  Tweedie}}]{Donnellan:2007xr}
\bibinfo{author}{\bibfnamefont{M.~A.} \bibnamefont{Donnellan}},
  \bibinfo{author}{\bibfnamefont{J.}~\bibnamefont{Flynn}},
  \bibinfo{author}{\bibfnamefont{A.}~\bibnamefont{Juttner}},
  \bibinfo{author}{\bibfnamefont{C.~T.} \bibnamefont{Sachrajda}},
  \bibinfo{author}{\bibfnamefont{D.}~\bibnamefont{Antonio}},
  \bibinfo{author}{\bibfnamefont{P.~A.} \bibnamefont{Boyle}},
  \bibinfo{author}{\bibfnamefont{C.}~\bibnamefont{Maynard}},
  \bibinfo{author}{\bibfnamefont{B.}~\bibnamefont{Pendleton}},
  \bibnamefont{and} \bibinfo{author}{\bibfnamefont{R.}~\bibnamefont{Tweedie}},
  \bibinfo{journal}{PoS} \textbf{\bibinfo{volume}{LATTICE2007}},
  \bibinfo{pages}{369} (\bibinfo{year}{2007}), \eprint{0710.0869}.

\bibitem[{\citenamefont{Pineda}(2001)}]{Pineda:2001zq}
\bibinfo{author}{\bibfnamefont{A.}~\bibnamefont{Pineda}},
  \bibinfo{journal}{JHEP} \textbf{\bibinfo{volume}{06}}, \bibinfo{pages}{022}
  (\bibinfo{year}{2001}), \eprint{hep-ph/0105008}.

\bibitem[{\citenamefont{Bali et~al.}(2013)\citenamefont{Bali, Bauer, Pineda,
  and Torrero}}]{Bali:2013pla}
\bibinfo{author}{\bibfnamefont{G.~S.} \bibnamefont{Bali}},
  \bibinfo{author}{\bibfnamefont{C.}~\bibnamefont{Bauer}},
  \bibinfo{author}{\bibfnamefont{A.}~\bibnamefont{Pineda}}, \bibnamefont{and}
  \bibinfo{author}{\bibfnamefont{C.}~\bibnamefont{Torrero}},
  \bibinfo{journal}{Phys. Rev. D} \textbf{\bibinfo{volume}{87}},
  \bibinfo{pages}{094517} (\bibinfo{year}{2013}), \eprint{1303.3279}.

\bibitem[{\citenamefont{Regge}(1959)}]{Regge:1959mz}
\bibinfo{author}{\bibfnamefont{T.}~\bibnamefont{Regge}},
  \bibinfo{journal}{Nuovo Cim.} \textbf{\bibinfo{volume}{14}},
  \bibinfo{pages}{951} (\bibinfo{year}{1959}).

\bibitem[{\citenamefont{Avkhadiev et~al.}(2023)\citenamefont{Avkhadiev,
  Shanahan, Wagman, and Zhao}}]{Avkhadiev:2023poz}
\bibinfo{author}{\bibfnamefont{A.}~\bibnamefont{Avkhadiev}},
  \bibinfo{author}{\bibfnamefont{P.~E.} \bibnamefont{Shanahan}},
  \bibinfo{author}{\bibfnamefont{M.~L.} \bibnamefont{Wagman}},
  \bibnamefont{and} \bibinfo{author}{\bibfnamefont{Y.}~\bibnamefont{Zhao}},
  \bibinfo{journal}{Phys. Rev. D} \textbf{\bibinfo{volume}{108}},
  \bibinfo{pages}{114505} (\bibinfo{year}{2023}), \eprint{2307.12359}.

\bibitem[{\citenamefont{Lin and Zhang}(2019)}]{Lin:2019ocg}
\bibinfo{author}{\bibfnamefont{H.-W.} \bibnamefont{Lin}} \bibnamefont{and}
  \bibinfo{author}{\bibfnamefont{R.}~\bibnamefont{Zhang}},
  \bibinfo{journal}{Phys. Rev. D} \textbf{\bibinfo{volume}{100}},
  \bibinfo{pages}{074502} (\bibinfo{year}{2019}).

\bibitem[{\citenamefont{Braun et~al.}(2021)\citenamefont{Braun, Manashov, Moch,
  and Schoenleber}}]{Braun:2021grd}
\bibinfo{author}{\bibfnamefont{V.~M.} \bibnamefont{Braun}},
  \bibinfo{author}{\bibfnamefont{A.~N.} \bibnamefont{Manashov}},
  \bibinfo{author}{\bibfnamefont{S.}~\bibnamefont{Moch}}, \bibnamefont{and}
  \bibinfo{author}{\bibfnamefont{J.}~\bibnamefont{Schoenleber}},
  \bibinfo{journal}{Phys. Rev. D} \textbf{\bibinfo{volume}{104}},
  \bibinfo{pages}{094007} (\bibinfo{year}{2021}), \eprint{2106.01437}.

\end{thebibliography}
\end{document}